\documentclass[12pt,a4paper]{report}
\usepackage{latexsym,amssymb,amsthm,amsmath}
\usepackage[margin=1in,noheadfoot]{geometry}
\usepackage{slashed}
\usepackage{setspace}
\usepackage{graphics,graphicx}

\begin{document}

\pagenumbering{roman}

\begin{titlepage}
\begin{center}
\LARGE \textbf{Diquark-Antidiquark Interpretation of Mesons}
\end{center}
\vspace{0.8in}

\begin{figure}[h]
    \centering
        \includegraphics[scale=.8]{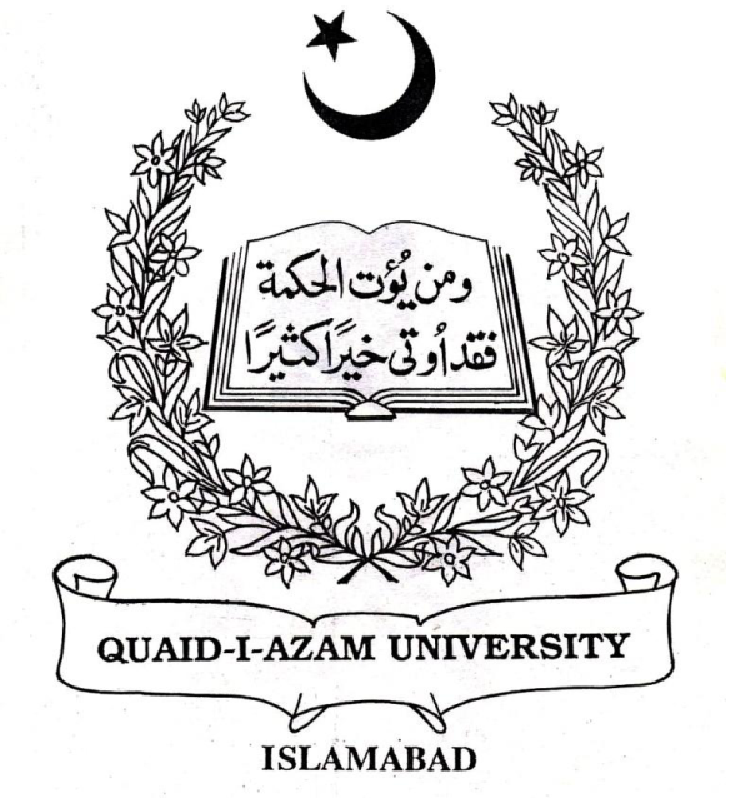}
    \label{fig:qau}
\end{figure}
\vspace{.4in}
\begin{center}
\Large\textbf{Abdur Rehman}\vspace{0.3in}\\ \textbf{Department of Physics}\\ \&\\ \textbf{National Centre for Physics}\\
\textbf{Quaid-e-Azam University}\\ \textbf{Islamabad, Pakistan} \\
\textbf{January, 2011.}
\end{center}

\pagebreak \pagestyle{empty}
\begin{center}
\large \textbf{This work is submitted as a dissertation in partial
fulfillment of the requirement for the degree of}
\end{center}
\vspace{0.3in}
\begin{center}
\LARGE \textbf{MASTER OF PHILOSOPHY\\ IN PHYSICS}
\end{center}
\vspace{0.3in}
\begin{figure}[h]
    \centering
        \includegraphics[scale=.8]{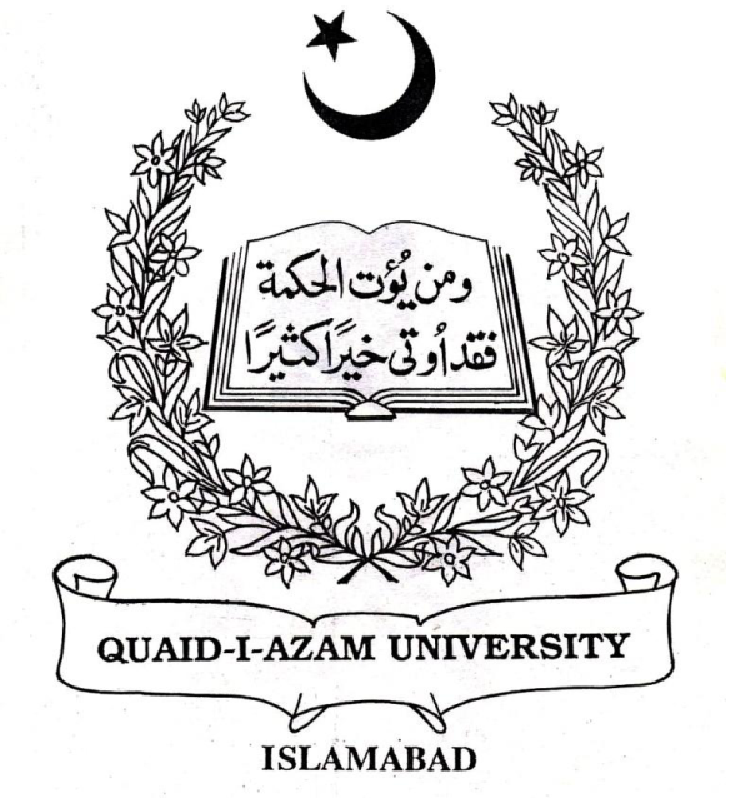}
    \label{fig:qau}
\end{figure}
\vspace{0.3in}
\begin{center}
\Large \textbf{Department of Physics}\\ \&\\ \textbf{National Centre for Physics}\\
\textbf{Quaid-e-Azam University}\\ \textbf{Islamabad, Pakistan} \\
\textbf{January, 2011.}
\end{center}

\pagebreak \pagestyle{empty}
\begin{center}
\LARGE \textbf{Certificate}
\end{center}

\begin{flushleft}
\large Certified that the work contained in this dissertation was
carried out by Mr. Abdur Rehman under my supervision.
\end{flushleft}

\vspace{1.5in}
\begin{flushright}
\large \textbf{Prof. Riazuddin}\\
Supervisor,\\
National Centre for Physics,\\
Quaid-i-Azam University, Islamabad\\
\&\\
CAMP-NUST, Islamabad, Pakistan.
\end{flushright}
\vspace{0.5in}
\begin{flushleft}
\large Submitted through
\end{flushleft}
\vspace{1in}
\begin{flushleft}
\large \textbf{Prof. S. K. Hasanain}\\
Chairman,\\
Department of Physics,\\
Quaid-i-Azam University,\\
Islamabad, Pakistan.
\end{flushleft}
\pagebreak \pagestyle{empty}

\begin{center}
\ \\
\ \\
\ \\
\ \\
\ \\
\ \\
\ \\
\ \\
\LARGE \textbf{\texttt{\textbf\emph{{To My Loving Ammi jee and Abbu jee }}}} \\

\end{center}

\end{titlepage}

\tableofcontents \listoffigures \listoftables

\chapter*{Acknowledgements}

All praise to Almighty Allah, the most Merciful, and Benevolent and
His beloved Prophet Muhammad (P.B.U.H). I want to express my
enormous gratitude to my supervisor, Prof. Riazuddin, for his
permanent encouragement to "work hard" and for wonderful example of
a persistent and motivated leader. I want specifically thank to
Muhammad Jamil Aslam for his guidance. Without his tremendous
influence, my research work would not be done. I am very thankful to
Prof. Fayyazuddin for many informative discussions and comments on
my work. I also want to thank Prof. Pervez Hoodbhoy and Dr. Hafiz
Horani who also taught me High Energy Physics courses with deep
theoretical and experimental concepts which provide me a great help
during my research work. In the last, I want to thank all my High
Energy Theory Group fellows at National Centre for Physics for both
their scientific and personal inputs in my life: Ishtiaq Ahmed, Ali
Paracha, Junaid, Aqeel, M. Zubair, Saadi Ishaq, Shahin Iqbal, Faisal
Munir, Khush Jan and people from other Institutions: Muhammad
Sadique Inam, Imran Shaukat, Shehbaz. I want to thank everybody who
is around me for the last few years for the constant support and
help and makes the campus life beautiful. This work was supported by
the National Centre for Physics, Islamabad. I would like to
appreciate the facilities provided by the centre. I am thankful to
Dr. Hamid Saleem for his wise input. And most of all I am grateful
to my parents, who supported me in all possible ways, and kept
inspiring me to work further when I was tired.

\vspace{0.5in}
\begin{flushright}
{\large \textbf{Abdur Rehman}}
\end{flushright}

\begin{abstract}
We study the spectroscopy of the states which defy conventional
$c\bar{c}$ charmonium and $b\bar{b}$ bottomonium interpretation
respectively, and are termed as exotic states. In August $2003$ a
state $X(3872)$, was discovered [K. Abe\textit{\ et al.} (Belle
Collaboration), Phys.\ Rev.\ Lett.\ \textbf{ 91}, 262001 (2003)
[arXiv:hep-ex/0309032]] and in December $2007$ a state $ Y(10890),$
was reported [K.~F.~Chen \textit{et. al.} [Belle Collaboration],
Phys.\ Rev.\ Lett.\ \textbf{100}, 112001 (2008) [arXiv:0710.2577].\
One possible interpretation of such exotic state is that they are
tetraquark (diquark-antiquark) states. Applying knowledge of
non-relativistic constituent quark model, we calculate the spectrum
of hidden charm and bottom exotic mesons within diquark-antidiquark
model. We investigate that $ X(3872)$ is $1^{++}$ state of the kind
$X_{\left[ cq\right] }=([cq]_{S=1}[
\bar{c}\bar{q}]_{S=0})_{S\mathrm{-wave}}$, and $Y(10890)$ is
$1^{--}$ state of the kind
$Y_{[bq]}=([bq]_{S=0}[\bar{b}\bar{q}]_{S=0})_{\mathrm{P-wave}}$ and
calculate the decay modes of these exotic states, further supporting
$ X(3872)$ \ and $Y(10890)$ as tetraquark (diquark-antidiquark)
states and resolve the puzzling features of the data. We study the
radiative decays of these states, using the idea of Vector Meson
Dominance (VMD), which we hope will increase an insight about these
tetraquark states.
\end{abstract}

\pagenumbering{arabic}

\chapter{Introduction}

In Heavy Quark physics one of the interesting problems is to
determine the properties of newly discovered particles. It becomes
very appealing when these particles do not fit into the existing
models. Quark model provides a convenient framework in the
classification of hadrons. Most experimentally observed hadronic
states fit in it nicely. The states which are beyond \ the quark
model are termed as \textbf{exotic}. To interpret these particles
many new hypothesis are created. Non-quark model mesons include:

1. \textbf{exotic mesons}, which defy conventional $q\bar{q}$
interpretation;

2. \textbf{glueballs or gluonium}, which have no valence quarks at
all;

3. \textbf{tetraquarks}, there are two configurations in tetraquark
picture, molecular models and Diquark-antiquark model;

4.\textbf{\ hybrid mesons}, which contain a valence quark-antiquark
pair and one or more gluons.

The exotic states in the charmonium spectrum have been found \
experimently, few of them are labelled as $X$, $Y$ and $Z$ states.
In $2003$ a particle temporarily called $X(3872)$, was discovered by
the Belle experiment in Japan \cite{16}. These exotic states refuse
to obey conventional $c\bar{c}$ charmonium interpretation
\cite{28,29}. So, many theories came: molecular models \cite{25,26},
more general 4-quark interpretations including a diquark-antiquark
model \cite{17}-\cite{21}, hybrid models \cite{27} etc. To explain
the nature of the $X\left( 3872\right) ,$ it was suggested as a
tetraquark candidate. The name $X$ is a temporary name, indicating
that there are still some questions about its properties which need
to be tested. The number in the parenthesis is the mass of the
particle in MeV. There are two configurations in\textbf{\
tetraquark} picture. In the first
configuration, binding each quark to an antiquark $\left[ q_{\alpha }\bar{q}%
^{\alpha }\right] $ and allowing interaction between the two color
neutral
pairs $\left[ q_{\alpha }\bar{q}^{\alpha }\right] \left[ q_{\beta }\bar{q}%
^{\beta }\right] $. This is what we call the molecular model. In
particular
the $X\left( 3872\right) $ happens to have a mass very close to the $D\bar{D}%
^{\ast }$ threshold. The binding energy left for the $X\left(
3872\right) $ is consistent with $E\thicksim 0.25\pm $ $0.40$ MeV,
thus making this state very large in size: order of ten times bigger
than the typical range of strong interactions. These questions apply
to other near-to-threshold hypothetical molecules and have induced
thinking to find alternative explanations for the $X\left(
3872\right) $ and its relatives. In the second configuration,
binding the two quarks in a colored configuration called
diquark $\left[ qq\right] _{\alpha }$, with antidiquark $\left[ \bar{q}\bar{q%
}\right] ^{\alpha }$. This configuration is what is called \textbf{%
diquark-antiquark model }where diquark is a fundamental object. The
$X(3872)$ is a $[cq]_{S=1}[\bar{c}\bar{q}]_{S=0}$ tetraquark. The
work of this dissertation is to discuss the lowest lying exotic
meson $X\left( 3872\right) $ and higher mass exotic state
$Y_{b}(10890)$ in the frame work of \textbf{diquark-antidiquark
model.}

In the diquark-antidiquark model the mass spectra are computed as in
the non-relativistic constituent quark model. In the constituent
quark model hadron masses are described by an effective Hamiltonian
that takes as input the constituent quark masses and the spin-spin
couplings between quarks. By extending this approach to
diquark-antidiquark bound states it is possible
to predict tetraquark mass spectra. The mass spectrum of tetraquarks $[%
\acute{q}q][\overline{\acute{q}}\bar{q}]$ with $q=u$, $d$ and
$\acute{q}=c$, $b$ neutral states can be described in terms of the
constituent diquark masses, $m_{\mathcal{Q}}$, spin-spin
interactions inside the single diquark, spin-spin interaction
between quark and antiquark belonging to two diquarks, spin-orbit,
and purely orbital term \cite{17}.

In the sceond chapter, we briefly discuss the Quark Model, hadron
spectroscopy and the\ concept of tetraquark. As\textbf{\ }diquark is
the fundamental object in the \textbf{diquark-antidiquark model}, a
complete section is devoted to understand its characteristics,
especially parity, color etc.

In chapter 3, we give a formulism of diquark-antiquark model. The
exotic state$\ X\left( 3872\right) $ is the focus of study in this
chapter. We calculate the spectrum of hidden charm states using this
model, which automatically shows that the $1^{++}$ state is the
$X\left( 3872\right) .$ In the last section we discuss the concept
of isospin symmetry breaking which helps to understand the finer
structure of the $X\left( 3872\right) $ and finally we calculate the
decay widths of $X\left( 3872\right) $. We calculate the radiative
decay widths of $X\left( 3872\right) $ by exploiting the idea of
Vector Meson Dominance (VMD).

In December $2007$, the Belle collaboration working at the KEKB
$e^{+}e^{-}$
collider in Tsukuba, Japan, reported the first observation of the processes $%
e^{+}e^{-}\rightarrow Y_{[bq]}\rightarrow \Upsilon (1S,2S)\;\pi
^{+}\pi ^{-}$ near the peak of the $\Upsilon (5S)$ resonance at the
center-of-mass energy of about $10.87$ $GeV$ \cite{41}. In the
conventional Quarkonium theory, there is no place for such a nearby
additional $b\bar{b}$ resonance having the quantum numbers of
$\Upsilon (5S).$ An important issue is whether the puzzling events
seen by Belle stem from the decays of the $\Upsilon (5S)$, or from
another particle $Y_{b}$ having a mass close enough to the mass of
the $\Upsilon (5S).$ The puzzling features of these data are that,
if interpreted in terms of the processes $e^{+}e^{-}\rightarrow
\Upsilon (5S)\rightarrow \Upsilon (1S)\;\pi ^{+}\pi ^{-},\Upsilon
(2S)\;\pi ^{+}\pi ^{-}$, the rates are anomalously larger than the
expectations from scaling the comparable $\Upsilon (4S)$ decays to
those of $\Upsilon (5S)$.

In the last chapter, we modify the formulism of diquark-antiquark
model to calculate the spectrum of hidden bottom states for $L=1$.
We are able to
show that $Y_{b}$ is $J^{PC}=1^{--}$ state, with $Y_{[bq]}=([bq]_{S=0}[\bar{b%
}\bar{q}]_{S=0})_{\mathrm{P-wave}},$ with the value $M_{Y_{[bq]}}^{(1)}$ $($%
for $q=u,d)$ equal to $10890$ MeV We identify this with the mass of the $%
Y_{b}$ from Belle \cite{37}, apart from the $\Upsilon (5S)$ and
$\Upsilon (6S)$ resonances. We calculated the leptonic, hadronic and
radiative decay widths of the $Y_{b}$ that may solve the puzzling
features of the data.

\chapter{THE QUARK MODEL AND BEYOND}

\section{Quarks}

\subsection{An Overview}

A quark is an elementary particle and a fundamental constituent of
matter. Quarks combine to form composite particles called hadrons.
Due to a phenomenon known as color confinement, quarks are never
found in isolation; they can only be found within hadrons. For this
reason, much of what is known about quarks has been drawn from
observations of the hadrons themselves \cite{1}. Quarks possess a
property called color charge. There are three types of color charge,
arbitrarily labeled blue, green, and red. Each of them is
complemented by an anticolor---antiblue, antigreen, and antired.
Every quark carries a color, while every antiquark carries an
anticolor. The theory that describes strong interactions is called
Quantum Chromodynamics (QCD).

The quarks which determine the quantum numbers of hadrons are called
valence quarks; apart from these, any hadron may contain an
indefinite number of virtual (or sea) quarks, antiquarks, and gluons
which do not influence its quantum numbers \cite{2}. There are two
families of hadrons: baryons, with three valence quarks, and mesons,
with a valence quark and an antiquark. The existence of "exotic"\
hadrons with more valence quarks, such as tetraquarks (qq\={q}\={q})
and pentaquarks (qqqq\={q}), have been conjectured but not proven
\cite{3}.

\begin{figure}[here]
\centering
    \includegraphics[scale=.6]{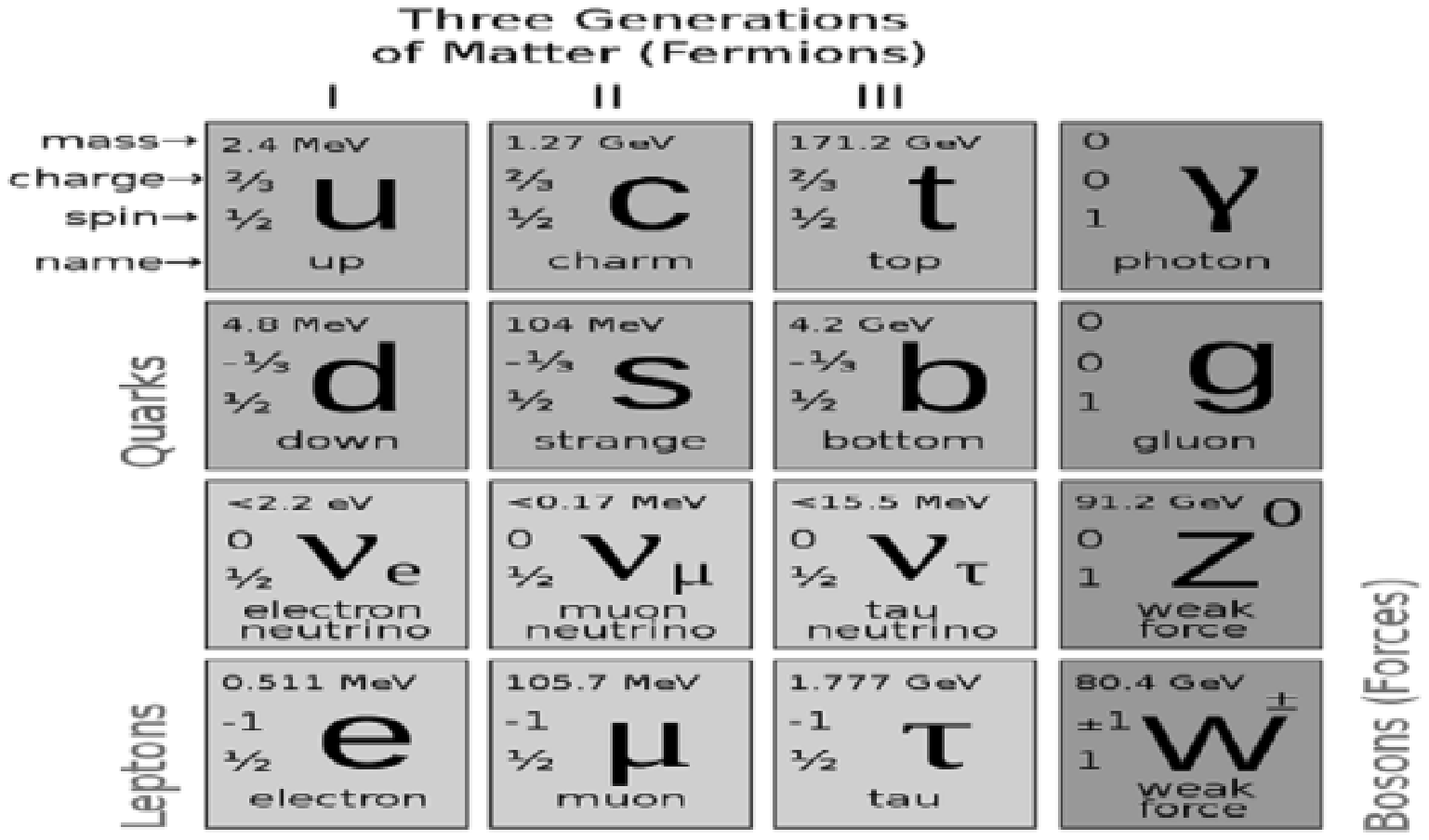}
    \caption{Six of the
particles in the Standard Model are quarks. Each of the first three
columns forms a generation of matter.}
\end{figure}

In modern particle physics, local gauge symmetries---a kind of
symmetry group---determine interactions between particles. Color
SU(3) (commonly abbreviated to SU$_{c}$(3)) is the gauge symmetry
that generated by three color charges which a quark carry and is the
defining symmetry for QCD. The requirement that SU$_{c}$(3) should
be local, i.e its transformations be allowed to vary with space and
time---determines the properties of the strong interaction, in
particular the existence of eight gluons to act as its force
carriers \cite{4}.

\subsubsection{Properties}

The Table 2.1 summarizes the key properties of the six quarks.
Flavor quantum numbers ( isospin $(I_{z})$, charmness $(C)$,
strangeness $(S$, not to be confused with spin$)$, topness $(T)$,
and bottomness $(B)$) are assigned to certain quark flavors, and
denote qualities of quark-based systems and hadrons. The baryon
number ($\acute{B}$) is $+1/3$ for all quarks, as baryons are made
of three quarks. For antiquarks, the electric charge $(Q)$ and all
flavor quantum numbers$(B,I_{z},C,S,T,$ $and$ $B)$ are of opposite
sign. Quarks are strongly interacting fermions with half integer
spin. Quarks have positive intrinsic parity, and are spin 1/2
particles. There are additive flavor quantum numbers for three
generations of quarks. Antiquarks have the opposite flavor sign. The
charge $Q$ and these quantum numbers are related through the
Gell-Mann-Nishijima relation:

\begin{equation}
Q=I_{z}+\frac{\acute{B}+S+C+B+T}{2}
\end{equation}%
where $\acute{B}=1/3,$ is the baryon number for each quark \cite{5}.

\begin{table}[h]
\caption{Properties of quarks}
\begin{center}
$%
\begin{tabular}{||c||c||c||c||c||c||c||c||c||}
\hline\hline $Quark\ \backslash \ \Pr operties$ & $Q$ & $I$ &
$I_{z}$ & $S$ & $C$ & $B$ & $T$ & $\acute{B}$ \\ \hline\hline $d$ &
$-\frac{1}{3}$ & $\frac{1}{2}$ & $-\frac{1}{2}$ & $0$ & $0$ & $0$ &
$0$ & $\frac{1}{3}$ \\ \hline\hline $u$ & $+\frac{2}{3}$ &
$\frac{1}{2}$ & $+\frac{1}{2}$ & $0$ & $0$ & $0$ & $0$ &
$\frac{1}{3}$ \\ \hline\hline
$s$ & $-\frac{1}{3}$ & $0$ & $0$ & $-1$ & $0$ & $0$ & $0$ & $\frac{1}{3}$ \\
\hline\hline
$c$ & $+\frac{2}{3}$ & $0$ & $0$ & $0$ & $+1$ & $0$ & $0$ & $\frac{1}{3}$ \\
\hline\hline
$b$ & $-\frac{1}{3}$ & $0$ & $0$ & $0$ & $0$ & $-1$ & $0$ & $\frac{1}{3}$ \\
\hline\hline
$t$ & $+\frac{2}{3}$ & $0$ & $0$ & $0$ & $0$ & $0$ & $+1$ & $\frac{1}{3}$ \\
\hline\hline
\end{tabular}%
$%
\end{center}
\end{table}
$\qquad \qquad \qquad \qquad \qquad \qquad \qquad \qquad \qquad
\qquad \qquad \qquad $

\section{Quark Model}

The quark model is a classification scheme for hadrons in term of
their valence quarks.

\subsection{Hadrons}

According to the quark model, the properties of hadrons are
primarily determined by their valence quarks. Although quarks also
carry color charge, hadrons must have zero color charge because of a
phenomena called color confinement. That is, hadrons must be
\textquotedblright colorless\textquotedblright\ or
\textquotedblright white\textquotedblright . There are two ways to
accomplish this: three quarks of different colors, or a quark of one
color and an antiquark carrying the corresponding anticolor. Hadrons
based on the former are called baryons (half odd integer spin), and
those based on the latter are called mesons (integer spin). This is
possible because of the remarkable property that a singlet exists in
$3\otimes
3\otimes 3=1\oplus 8\oplus 8\oplus 10$ as well as in $3\otimes \bar{3}%
=1\oplus 8$. All hadrons are labelled by quantum numbers. One set
comes from the Poincar\'{e} symmetry- $J^{PC}$, where $J$, $P$ \ and
\thinspace $C$ \ stand for the total angular momentum, Parity, and
Charge-symmetry respectively. Hadrons with same $J^{P}$ (lowest
lying as well as higher mass states) are distinguished from each
other by some internal quantum numbers.\ These are flavor quantum
numbers such as the isospin, strangeness, charm, and so on. All
quarks carry an additive, conserved quantum number called a baryon
number, which is +1/3 for quarks and -1/3 for antiquarks. This means
that baryons (groups of three quarks) have B = 1 while mesons have B
= 0. Hadrons have excited states known as resonances. Each ground
state hadron may have several excited states. Resonances decay
extreme quickly (within about $10^{-24}$ seconds) via the strong
nuclear force \cite{3}.

\subsection{ Baryons}
\begin{figure}[here]
\centering
    \includegraphics[scale=.5]{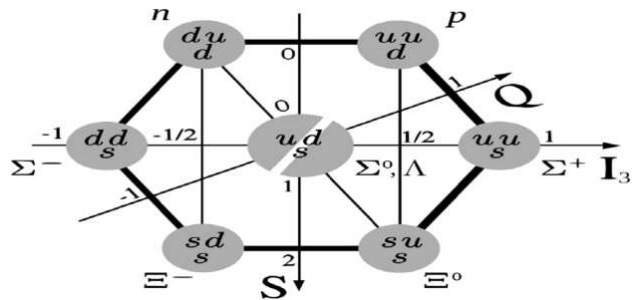}
    \caption{Combinations
of three $u,d$ or $s$ quarks forming baryons with a spin $1/2$ form
the $uds$ baryon octet.}
\end{figure}
\begin{figure}[here]
\centering
    \includegraphics[scale=.5]{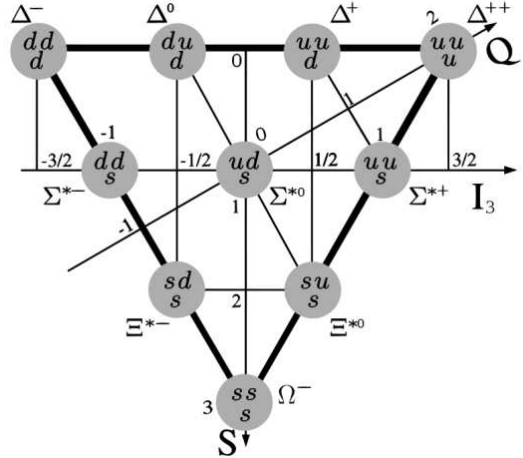}
    \caption{Combinations
of three $u,d$ or $s$ quarks forming baryons with a spin $3/2$ from
the $uds$ baryon decuplet.}
\end{figure}

\subsection{Exotic Baryons}

Exotic baryons are hypothetical composite particles which are bound
states of 3 quarks and additional elementary particles. The
additional particles may include quarks, antiquarks or gluons. One
such exotic baryon is the pentaquark, which consists of four quarks
and an antiquark (\'{B}= 1), but their existence is not generally
accepted. Theoretically, heptaquarks (5 quarks, 2 antiquarks),
nonaquarks (6 quarks, 3 antiquarks), etc. could also exist. Another
exotic baryon which consists only of quarks is the H dibaryon, which
consists of two up quarks, two down quarks and two strange quarks.
Unlike the pentaquark, this particle might be long lived or even
stable. There have been unconfirmed claims of detections of
pentaquarks and dibaryons \cite{7}.

\subsection{Mesons}

The main difference between mesons and baryons is that mesons are
bosons (which\ obey Bose-Einstein statistics) while baryons are
fermions (which\ obey Fermi-Dirac statistics). Since mesons are
composed of quarks, they participate in both the weak and strong
interactions. Mesons with net electric charge also participate in
the electromagnetic interaction. They are classified according to
their quark content, total angular momentum, parity, and various
other properties such as $C-$parity and $G-$parity. They are also
typically less massive than baryons, meaning that they are more
easily produced in experiments, and exhibit higher energy phenomena
sooner than baryons would. For example, the charm quark was first
seen in the J/Psi
meson ($J/\Psi $) in 1974, and the bottom quark in the upsilon meson ($%
\Upsilon $) in $1977$ \cite{8}.

\subsubsection{Classification Of Mesons}

The mesons are classified in $J^{PC}$ multiplets. The allowed
quantum numbers for $L$ smaller than $3$ are given in table.
Interestingly, for $J$ smaller than $3$, all allowed $J^{PC}$\
except $2^{--}$ \ have been observed \cite{3}.

\begin{table}[h]
\caption{Classification of mesons}
\begin{center}
$%
\begin{tabular}{||l||l||l||l||l||l||l||l||l||}
\hline\hline
$L$ & $S$ & $J^{PC}$ & $L$ & $S$ & $J^{PC}$ & $L$ & $S$ & $J^{PC}$ \\
\hline\hline
$0$ & $0$ & $0^{-+}$ & $1$ & $0$ & $1^{+-}$ & $2$ & $0$ & $2^{-+}$ \\
\hline\hline
$0$ & $1$ & $1^{--}$ & $1$ & $1$ & $0^{++}$ & $2$ & $1$ & $1^{--}$ \\
\hline\hline &  &  & $1$ & $1$ & $1^{++}$ & $2$ & $1$ & $2^{--}$ \\
\hline\hline &  &  & $1$ & $1$ & $2^{++}$ & $2$ & $1$ & $3^{--}$ \\
\hline\hline
\end{tabular}%
$%
\end{center}
\end{table}

Particle physicists are most interested in mesons with no orbital
angular
momentum $(L=0)$, therefore the two groups of mesons most studied are the $%
S=1;$ $L=0$ and $S=0;$ $L=0$, which corresponds to $J=1$ and $J=0$,
although they are not the only ones. It is also possible to obtain
$J=1$ particles from $S=0$ and $L=1$. How to distinguish between the
$S=1,$ $L=0$ and $S=0,$ $L=1$ mesons is an active area of research
in meson spectroscopy.

For lighter up, down, and strange quarks the suitable mathematical
group is\ $SU(3)$. The quarks lie in the fundamental representation,
$3$ (called the triplet) of flavor $SU(3)$. The antiquarks lie in
the complex conjugate representation $\bar{3}$. The nine states
(nonet) made out of a pair can be decomposed into the trivial
representation, $1$ (called the singlet), and the adjoint
representation, $8$ (called the octet). The notation for this
decomposition is $3\otimes \bar{3}=8\oplus 1$. There are
generalizations to
larger number of flavors. Charm quark is icluded by extending$\ SU(3)$\ to $%
SU(4)$ \cite{3}.

\subsubsection{Types Of Mesons}

The rules for classification\ are presented below, in Table 2.3 for
simplicity \cite{11}.

\begin{table}[h]
\caption{Types of mesons}
\begin{center}
\begin{tabular}{||l||l||l||l||l||l||}
\hline\hline $L$ & $J^{PC}$ & $Type$ & $L$ & $J^{PC}$ & $Type$ \\
\hline\hline $0$ & $0^{-+}$ & $pseudoscalars$ & $1$ & $0^{++}$ &
$scalars$ \\ \hline\hline $0$ & $1^{--}$ & $vectors$ & $1$ &
$1^{++},1^{+-}$ & $axial\text{ }vectors$
\\ \hline\hline
&  &  & $1$ & $2^{++}$ & $tensors$ \\ \hline\hline
\end{tabular}%
\end{center}
\end{table}

\begin{figure}[here]
\centering
    \includegraphics[scale=.5]{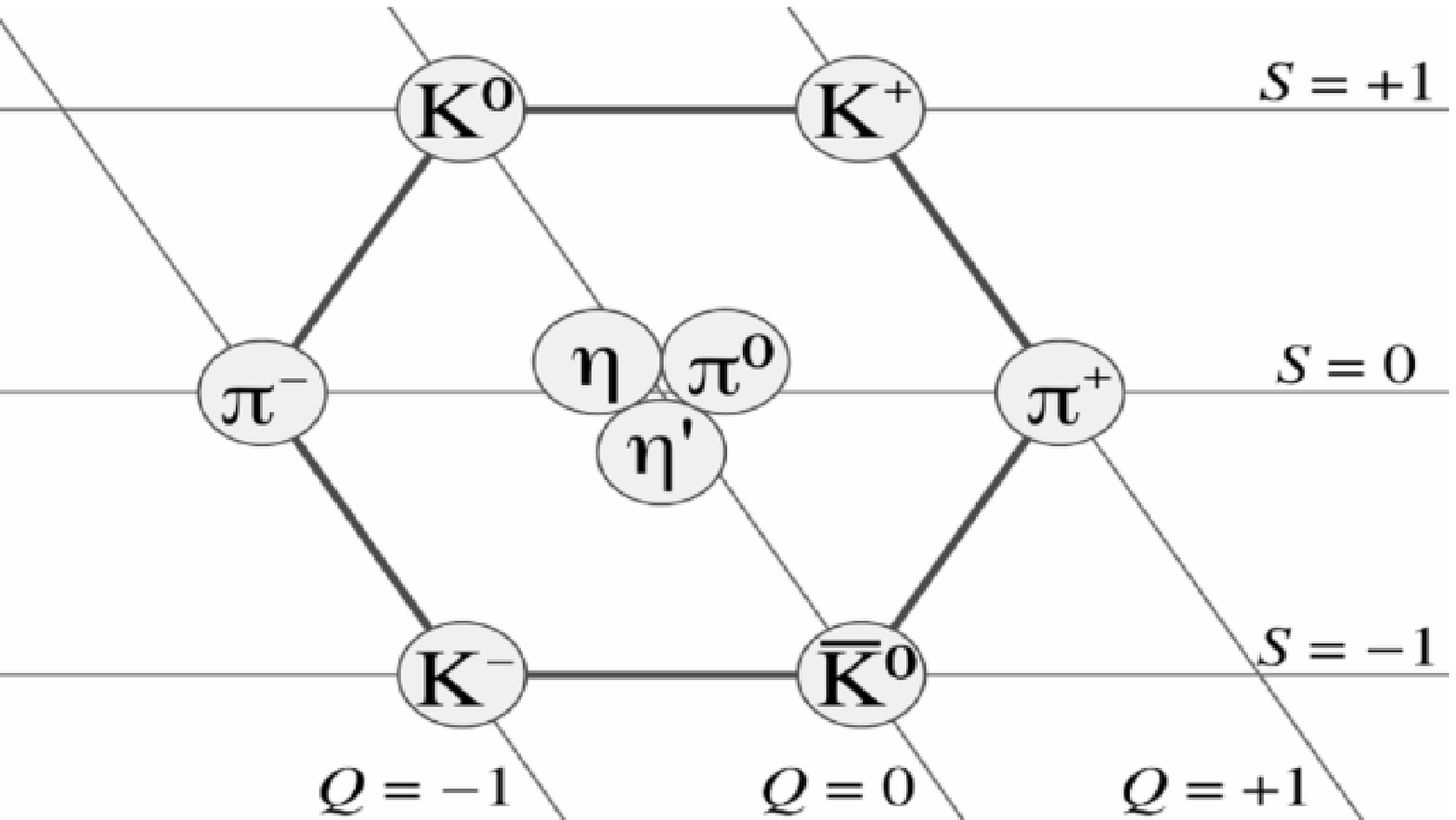}
    \caption{Combinations of one $u,$ $d$ or $
s $ quarks and one $u$, $d$ or $s$ antiquark in $J^{P}=0^{-}$
configuration form a nonet.}
\end{figure}

\begin{figure}[here]
\centering
    \includegraphics[scale=.5]{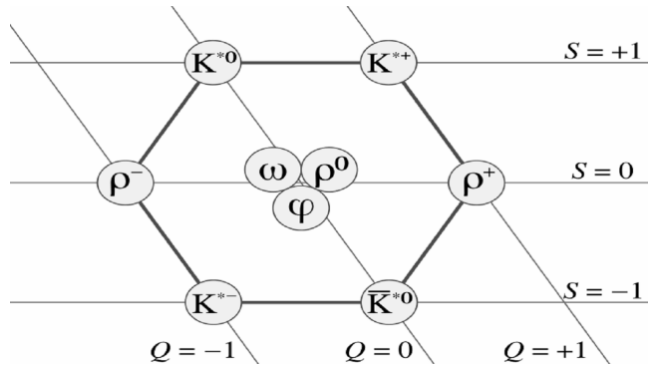}
    \caption{Combinations of one $u,$ $d$ or $
s $ quarks and one $u$, $d$ or $s$ antiquark in $J^{P}=1^{-}$
configuration form a nonet.}
\end{figure}

Flavorless mesons are mesons made of quark and antiquark of the same
flavor (all their flavor quantum numbers are zero. Flavorful mesons
are mesons made of pair of quark and antiquarks of different
flavors.

\section{Mesons And Symmetries}

Most of the symmetries in elementary-particle physics are
continuous. A typical example is the symmetry generated by rotations
around an axis, where the angle of rotation can assume any value
between $0$ and $2\pi $. In addition to continuous symmetries, there
are also discrete symmetries, for which the possible states assume
discrete values classified with the help of a few integers. In
elementary-particle physics there are three discrete symmetries of
basic importance: parity, charge conjugation and time-reversal.

Spin (quantum number $S$) is a vector quantity that represents the
\textquotedblright intrinsic\textquotedblright\ angular momentum of
a particle. Since\ quarks are fermi particles of spin $1/2$
$(S=1/2)$ and mesons are made of one quark and one antiquark, they
can be found in triplets and singlets spin states. The orbital
angular momentum (quantum number L), that comes in increments of
$1\hslash $, represent the angular moment due to quarks orbiting
around each other. The total angular momentum (quantum number J) of
a particle is the combination of intrinsic angular
momentum (spin) and orbital angular momentum. It can take any value from $%
J=|L-S|$ to $J=|L+S|$, in increments of $1$.

\subsection{C- and P- parity, and Isospin}

An important property of any quark state is the behavior of its
wave-funtion under certian transformations. One important
transformation is the replacement of particle with anti-particle,
called C-cojugation. An other one, called P-transformation, is the
switching the signs of all coordinates. Many states, but not all,
are eigenstates of these two transformations. This means that:
\begin{eqnarray*}
\Phi ^{\prime } &=&C\Phi =\lambda _{C}\Phi \\
\Phi ^{\prime } &=&P\Phi =\lambda _{P}\Phi
\end{eqnarray*}%
The former is possible for the states which are flavor neutral, e.g.
electrically neutral. These numbers, $\lambda _{C}$ and $\lambda
_{P}$, are called the $C$-parity and $P$- parity of the particular
quark state respectively. These transformations have one special
property:

\begin{eqnarray}
C^{2}\Phi &=&\lambda _{C}^{2}\Phi =\Phi  \label{1.2} \\
P^{2}\Phi &=&\lambda _{P}^{2}\Phi =\Phi  \label{1.3}
\end{eqnarray}%
It follows that the particular quark state may have $C$ \bigskip or $P-$%
parity either $+1$ or $-1$. For example, the $P-$parity of $\pi -$mesons is $%
-1$:

\[
P\pi ^{0}=-\pi ^{0},P\pi ^{+}=-\pi ^{+},P\pi ^{-}=-\pi ^{-}
\]%
Here $\pi $ stands for the $\pi -$meson wavefuntion. The $C-$parity
for $\pi ^{0}$ is $+1$, $C\pi ^{0}=+\pi ^{0};$ but $\pi ^{+}$ and
$\pi ^{-}$ do not have definite $C-$parity,\textit{\ i.e. }$C\pi
^{+}=+\pi ^{-},$ $C\pi ^{-}=+\pi ^{+}.$

For the system of a quark and an antiquark $(q\bar{q})_{L}$ [note
that $q$
and $\bar{q}$ have opposite intrinsic parities], one has%
\begin{equation}
P=(-1)\left( -1\right) ^{L},\text{ }C=\left( -1\right) ^{L+S}\text{\
\ \ \ } \label{1.4}
\end{equation}%
For states composite of integer spin bosons these formulae are
different. For example, for $\pi ^{+}\pi ^{-}$ system where pions
have zero spin and
same intrinsic parities.%
\begin{equation}
P=C=\left( -1\right) ^{L}  \label{1.5}
\end{equation}%
An important property of \ $C$ and $P-$parities is that they are\
conserved in the strong and electromagnetic interactions \cite{6}. A
generalization to $C-$parity is $G-$parity $G=\left( -1\right)
^{L+S+I}$ for mesons. The isospin of $u$ and $d$ quarks is equal to
$1/2$, with $u$ quark having
positive isospin projection $I$ $_{z}=1/2$, and $d$ quark with $I$ $%
_{z}=-1/2 $. All other quarks have zero isospin and same holds for
a\ state which is made up of these quarks. For example the isospin
of any charmonium state is zero. In particular, isospin becomes
crucial when we discuss the possible assignment for the $X\left(
3872\right) ,$ which is assumed to tetraquark state. Isospin $I$ is
another property of quark-antiquark system, which is conserved in
strong interactions and follow the same algebraic rules as the
regular spin $S$. Hadron with nearly same mass can be put into
isospin multiplets:
\begin{table}[h]
\caption{Isospin Multiplets}
\begin{center}
$%
\begin{tabular}{||l||l||l||}
\hline\hline $I=\frac{1}{2}$ & $\left(
\begin{tabular}{l}
$p$ \\
$n$%
\end{tabular}%
\right) ,$ $\ \left(
\begin{tabular}{l}
$K^{+}$ \\
$K^{0}$%
\end{tabular}%
\right) $ & $%
\begin{tabular}{l}
$I_{z}=+\frac{1}{2}$ \\
$\ \ =-\frac{1}{2}$%
\end{tabular}%
$ \\ \hline\hline $I=1$ & $\left(
\begin{tabular}{l}
$\Sigma ^{+}$ \\
$\Sigma ^{0}$ \\
$\Sigma ^{-}$%
\end{tabular}%
\right) ,$ \ $\left(
\begin{tabular}{l}
$\pi ^{+}$ \\
$\pi ^{0}$ \\
$\pi ^{-}$%
\end{tabular}%
\right) $ & $%
\begin{tabular}{l}
$I_{z}=1$ \\
$\ \ \ =0$ \\
$\ \ \ =-1$%
\end{tabular}%
$ \\ \hline\hline
\end{tabular}%
$%
\end{center}
\end{table}
\newline
Isospin is conserved in strong interaction and as such in that
limit, member of each multiplet will have the same mass. The small
difference then arises
due to electromagnetic\ interaction, which still conserves $I_{z}$, since $%
Q=I_{z}+isoscalar$ and charge conservation then implies $I_{z}$
conservtion
and/or $m_{u}$ $\neq m_{d}$. This is illustrated by the following decay.%
\[
\Sigma ^{0}|I=1,I_{3}=0\rangle \rightarrow \Lambda \gamma
|I=0,I_{3}=0\rangle
\]%
Isospin and its third component are not conserved in the weak
interaction,
as demonstrated in the decay:%
\[
\Lambda |0,0\rangle \rightarrow \pi ^{-}p|1,-1\rangle |\frac{1}{2},-\frac{1}{%
2}\rangle =\sqrt{\frac{1}{3}}|\frac{3}{2},\frac{1}{2}\rangle -\sqrt{\frac{2}{%
3}}|\frac{1}{2},-\frac{1}{2}\rangle
\]

\subsection{Isospin Symmetry}

Isospin symmetry, which is an exact symmetry as for as strong
interaction is concerned, is broken by electromagnetic interaction
and/or $m_{u}$ $\neq m_{d}$. Thus QCD Lagrangian has isospin
symmetry. If $m_{u}$ $=m_{d}=0$ the QCD Lagrangian has chiral
symmetry. Since $m_{u}\approx m_{d}<<\Lambda
_{QCD} $, the symmetry of the QCD lagrangian is broken when $m_{u}$ $%
=m_{d}\neq 0$.

\subsection{Isospin, Charge and Flavor Quantum Numbers}

The pion particle had three \textquotedblleft charged
states\textquotedblright , it was said to be of isospin $I=1$. Its
\textquotedblleft charged states\textquotedblright\ $\pi ^{+}$, $\pi
^{0}$,
and $\pi ^{-}$, corresponded to the isospin projections $I_{z}=+1,$ $%
I_{z}=0, $ and $I_{z}=-1$ respectively. Another example is the
\textquotedblleft rho particle\textquotedblright . Isospin
projections were
related to the up and down quark content of particles by the relation%
\begin{equation}
I_{z}=\frac{1}{2}[(n_{u}-n_{\bar{u}})-(n_{d}-n_{\bar{d}})]
\label{1.6}
\end{equation}%
where the n's are the number of up and down quarks and antiquarks.

It was noted that charge $(Q)$ was related to the isospin projection ($I$ $%
_{z}$), the baryon number $(B)$ and flavor quantum numbers $(S,C,\acute{B}%
,T)$ by the Gell-Mann--Nishijima formula. Strangeness flavor quantum
number $S$ (not to be confused with spin). Flavor quantum numbers of
composites are related to the number of strange, charm, bottom, and
top quarks and
antiquark according to the relations:%
\begin{eqnarray*}
S &=&-(n_{s}-n_{\bar{s}}) \\
C &=&+(n_{c}-n_{\bar{c}}) \\
\acute{B} &=&-(n_{b}-n_{\bar{b}}) \\
T &=&+(n_{t}-n_{\bar{t}})
\end{eqnarray*}
This implies that the Gell-Mann--Nishijima formula is equivalent to
the
expression of charge in terms of quark content \cite{10}.%
\begin{equation}
Q=\frac{2}{3}[(n_{u}-n_{\bar{u}})+(n_{c}-n_{\bar{c}})+(n_{t}-n_{\bar{t}})]-%
\frac{1}{3}[(n_{d}-n_{\bar{d}})+(n_{s}-n_{\bar{s}})+(n_{b}-n_{\bar{b}})]
\label{1.7}
\end{equation}

\section{Exotic Meson}

From the quantum numbers in Table 2.1-2.3, there are several
combinations which are missing:
\[
0^{+-},0^{--},1^{-+}\text{ \ \ and }2^{+-}
\]%
These are not possible for simple $q\bar{q}$ systems and are known
as "exotic" states. Beyond the simple quark model picture of mesons,
there are different frameworks suggested to accomodate\ these states
with exotic quantum numbers. Non-quark model mesons include:

1. glueballs or gluonium, which have no valence quarks at all;

2. tetraquarks, which have two valence quark-antiquark pairs; and

3. hybrid mesons, which contain a valence quark-antiquark pair and
one or more gluons.

All of these can be classfied as mesons, because they carry zero
baryon number. Of these, glueballs must be flavor singlets; that is,
have zero isospin, strangeness, charm, bottomness, and topness. Like
all particle states, they are specified by the quantum numbers
$J^{PC}$ and by the mass. One also specifies the isospin $I$ of the
meson. It is an old idea that the light scalar mesons a($980$) and
f($980$) may be 4-quark bound states. The idea was more or less
accepted in the mid-seventies but then it losts momentum, due to
contradictory results. If the lightest scalar mesons are
diquark-antidiquark composites as shown below, it is natural to
consider analogous states with one or more heavy constituents, to be
discussed in the following section.
\begin{figure}[here]
\centering
    \includegraphics[scale=.5]{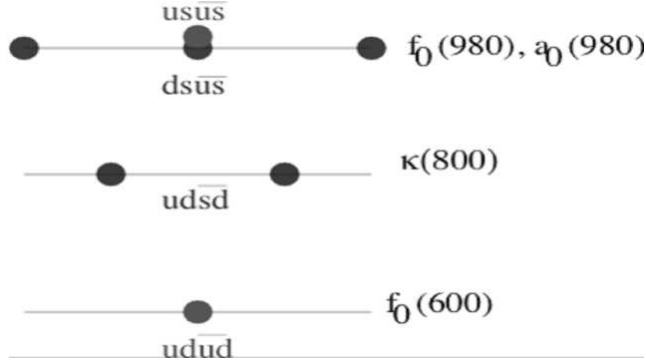}
    \caption{Identities and
classification of possible tetraquark mesons. First horizontal line
denotes I = 0 states, second, I = 1/2 and third one, I = 1. The
vertical axis is the mass.}
\end{figure}

\section{Diquarks: An Introduction}

The notion of the diquark usually means the system of two rather
tightly bounded quarks with a small size of$\ 0.1-0.3$ Fermi
\cite{15}. This section is devoted to diquarks and their role in
understanding exotics in QCD.\ Diquarks are not new, they are almost
as old as QCD. Gell-Mann mentions it prominently in his first paper
on quarks in $1964$ \cite{13}.

Baryons can be constructed from quarks by using the combinations $(qqq),$ $%
(qqqq\bar{q})$ etc, while mesons are made out of $\ (q\bar{q}),$ $(qq\bar{q}%
\bar{q}),$ etc. The lowest baryon configuration $(qqq)$ gives just
the representations $1,$ $8,$ and $10$ that have been observed,
while the lowest meson configuration $(q\bar{q})$ similarly gives
just $1$ and $8$.\

The constituents of the tetraquarks, diquarks and antidiquarks, have
well-defined properties, characterized by their color and
electromagnetic charges, spin and flavor quantum numbers. The
tetraquark hadrons are singlets in color (pictured white), and hence
they participate as physical states in scattering and decay
processes. This is not too dissimilar a situation from the
well-known mesons, which are color singlet (white) bound states of
the confined colored quarks and antiquarks.

We follow the suggestion by Jaffe and Wilczek $(1977)$ of having
diquark as building blocks \cite{2}. Diquark correlations in hadrons
suggest qualitative explanations for many of the puzzles of exotic
hadron spectroscopy. Operators that will create a diquark of any
(integer) spin and parity can be constructed from two quark fields
and insertions of the covariant derivative. We are interested in
potentially low energy configurations, so we omit the derivatives.
There are eight distinct diquark multiplets (in color$\times
$flavor$\times $spin), which are enumerated by
R. L.Jaffe. Since each quark is a color triplet, the pair can form a color $%
\bar{3}_{c}$, which is antisymmetric, or $6_{c}$, which is
symmetric. The same is true in SU(3)-flavor. The constructions look
more familiar if we represent one of the quarks by the charge
conjugate field: $qq\rightarrow \bar{q}_{c}q$, where
$\bar{q}_{c}=-iq^{T}\sigma ^{2}\gamma _{5}$. Then the
classification of diquark bilinears is analogous to the classification of $q%
\bar{q}$ bilinears. There are only two favored configurations. The
most attractive channel\ in QCD seems to be the color antitriplet,
flavor antisymmetric (which is the $\bar{3}_{f}$ for three light
flavors), spin singlet with even parity:
$[qq]^{\bar{3}_{c},\bar{3}_{f},0^{+}}.$ This
channel is favored by one gluon exchange and by instanton interactions \cite%
{40}. It will play the central role in the exotic drama.%
\begin{eqnarray}
&&|\{qq\}\bar{3}_{c}(A)\bar{3}_{f}(A)0^{+}(A)\rangle \text{ \ \ good diquarks%
}  \label{1.8} \\
&&|\{qq\}\bar{3}_{c}(A)6_{f}(S)1^{+}(S)\rangle \text{ \ \ bad
diquarks} \label{1.9}
\end{eqnarray}%
\ \ Both of these configurations are important in spectroscopy. Now
we can construct the operators for "good" scalar diquarks and "bad"
vector diquarks \cite{20}. Heavy-light diquarks can be the building
blocks of a rich spectrum of states which can accommodate some of
the newly observed
charmonium-like resonances not fitting a pure $c\bar{c}$ assignment.%
\begin{eqnarray}
\mathbf{Q}_{ia} &=&\epsilon _{ijk}\epsilon _{abc}(i\sigma
_{2})q^{jb}q^{kc}=\epsilon _{ijk}\epsilon
_{abc}(\bar{q}_{c}^{jb}\gamma
_{5}q^{kc})  \label{1.11(a)} \\
\mathbf{\acute{Q}}_{a}^{ij} &=&\epsilon _{abc}(\bar{q}_{c}^{jb}\vec{\gamma}%
q^{ic}+\bar{q}_{c}^{ib}\vec{\gamma}q^{jc})  \label{1.11(b)}
\end{eqnarray}%
Both represent positive parity, $0^{+}$ and $1^{+}$, states. We work
out the diquark masses in quark model, in which all residual quark
interactions are incorporated. Diquarks are, of course, colored
states, and therefore not physical. The good (scalar) and bad
(vector) diquarks configurations are our main interest.

\section{Tetraquark}

A tetraquark is a hypothetical meson composed of four valence
quarks. In principle, a tetraquark state may be allowed in Quantum
chromodynamics, the modern theory of strong interactions. Examining
the color algebra of the
system reveals there are two independent tetraquark singlet states: $%
3\otimes \bar{3}\otimes 3\otimes \bar{3}$ $=27\oplus 10\oplus
10\oplus 8\oplus 8\oplus 8\oplus 8\oplus 1\oplus 1$. But they can be
obtained in four different ways, depending on the intermediate color
states: the singlet scheme (or molecule), the octet scheme, the
triplet scheme and the sextet scheme.
\begin{figure}[here]
\centering
    \includegraphics[scale=.5]{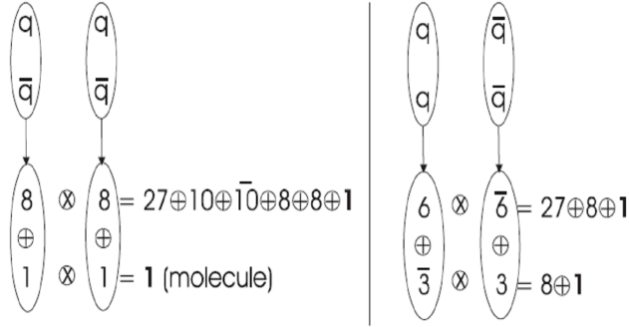}
    \caption{Singlets can be obtained
via four different schemes (only two are independent): octet,
singlet (or molecule), sextet and triplet.}
\end{figure}

 A diquark is either in symmetric color state $6_{c}$ or
color antisymmetric state $\bar{3}_{c}$. The antidiquark is either
in antisymmetric color state $\bar{6}_{c}$ or color symmetric state
$3_{c}$. Now
\begin{eqnarray*}
6_{c}\otimes \overline{6}_{c} &=&35_{c}\oplus 1_{c} \\
\overline{3}_{c}\otimes 3_{c} &=&8_{c}\oplus 1_{c} \\
\overline{3}_{c}\otimes \overline{6}_{c} &=&\overline{10}_{c}\oplus 8_{c} \\
6_{c}\otimes 3_{c} &=&10_{c}\oplus 8_{c}
\end{eqnarray*}%
Hence only $6_{c}\otimes \overline{6}_{c}$ and
$\overline{3}_{c}\otimes 3_{c} $ give color singlet state.

Now diquark are either in symmetric or antisymmetric in flavor:%
\begin{eqnarray*}
\left[ qq\right] &=&\frac{1}{\sqrt{2}}(q_{i}q_{j}-q_{j}q_{i})\text{ \ }i,%
\text{ }j=u\text{, }d,\text{ }s,\text{ }c \\
\left\{ qq\right\} &=&\frac{1}{\sqrt{2}}(q_{i}q_{j}+q_{j}q_{i})
\end{eqnarray*}%
For antisymmetric color state $\overline{3}_{c}$ or $3_{c}$, Pauli
principle requires, overall wave function of diquark or antidiquark
to be symmetric in flavor, in space and spin: (s, denote the spin of
diquark or antidiquark)
\begin{eqnarray*}
\left[ qq\right] _{L=0,\text{ }s=0}\text{ \ \ \ };\text{ }P
&=&1\text{ \ \ \
\ }\left[ qq\right] _{L=1,\text{ }s=1}\text{ \ \ \ };\text{ }P=-1 \\
\left\{ qq\right\} _{L=0,\text{ }s=1}\text{ \ \ \ };\text{ }P
&=&1\text{ \ \
\ \ }\left\{ qq\right\} _{L=1,\text{ }s=0}\text{ \ \ \ };\text{ }P=-1 \\
\left[ \bar{q}\bar{q}\right] _{L=0,\text{ }s=0}\text{ \ \ \ };\text{ }P &=&1%
\text{ \ \ \ \ }\left[ \bar{q}\bar{q}\right] _{L=1,\text{ }s=1}\text{ \ \ \ }%
;\text{ }P=-1 \\
\left\{ \bar{q}\bar{q}\right\} _{L=0,\text{ }s=1}\text{ \ \ \
};\text{ }P &=&1\text{ \ \ \ \ }\left\{ \bar{q}\bar{q}\right\}
_{L=1,\text{ }s=0}\text{ \ \ \ };\text{ }P=-1
\end{eqnarray*}%
We have a nonet of low lying scalar mesons $0^{+}$, composite of
tetraquark
viz%
\[
\left[ qq\right] _{L=0,\text{ }s=0}\ \left[ \bar{q}\bar{q}\right] _{L=0,%
\text{ }s=0}
\]%
As an example, we consider the following two tetraquark quarkonium
states, with $L=0$:

\[
\left( \left[ qq\right] _{L=0,\text{ }s=0}\left\{
\bar{q}\bar{q}\right\}
_{L=0,\text{ }s=1}\pm \left\{ qq\right\} _{L=0,\text{ }s=1}\left[ \bar{q}%
\bar{q}\right] _{L=0,\text{ }s=0}\right)
\]%
The two states have $J^{PC\text{ }}=1^{++}$ and $\ 1^{+-}.$ For $q=c$ the $%
J^{PC\text{ }}=1^{++}$ meson is identified with the state $X(3872)$.
This state was reported by the Belle in $2003$. It was suggested as
a tetraquark candidate. This state was discovered in the $J/\Psi $
$\pi ^{+}\pi ^{-\text{ }}$ distribution.

In $2004$ the $D(2632)$ state, seen in the SELEX experiment, was
suggested
as a possible tetraquark candidate with quark contents $[cd][\bar{d}\bar{s}]$%
. In $2009$ Fermilab announced that they have discovered a particle
temporarily called $Y(4140)$, which may also be a tetraquark with
quark contents $[cs][\bar{c}\bar{s}]$ \cite{12}.

\section{QCD, Quarkonia and Hadron Spectroscopy}

\subsection{The Important Physical Properties of QCD}

\subsubsection{Gluons}

The gluons, being mediators of strong interaction between quarks,
are vector particles and carry color; both of these properties are
supported by hadron spectroscopy.

\subsubsection{Color Confinement}

Confinement which implies that potential energy between color
charges increases linearly at large distances so that only color
singlet states exist, a property not yet established but find
support from lattice simulations and qualitative pictures. The
reasons for quark confinement are somewhat complicated; no analytic
proof exists that quantum chromodynamics should be confining, but
intuitively, confinement is due to the force-carrying gluons having
color charge.

\subsubsection{Asymptotic Freedom}

Asymptotic freedom which implies that the effective coupling constant $%
\alpha _{s}=\frac{g_{s}^{2}}{4\pi }$ decreases logarithmically at
short distances or high momentum transfers, a property which has a
rigorous theoretical basis. This is the basis for perturbative QCD
which is relevant
for processes involving large momentum transfers. we have%
\begin{equation}
\alpha _{s}(Q^{2})=\frac{2\pi }{(11-\frac{2}{3}n_{f})\ln \frac{Q^{2}}{%
\Lambda _{QCD}^{2}}}  \label{1.12}
\end{equation}%
the running of $\alpha _{s}(Q^{2})$ with $Q^{2}.\Lambda _{QCD}$ is
the QCD scale factor which effectively defines the energy scale at
which the running coupling constant attains its maximum value. For
$\frac{2}{3}n_{f}$ $<11$, it is clear that $\alpha _{s}(Q^{2})$\
decreases as $Q^{2}$ increases and approaches zero as
$Q^{2}\rightarrow \infty $ or $r\rightarrow 0$. This is known as the
asympotic freedom property of QCD \cite{6}.

\subsection{Quarkonia And Hadron Spectroscopy}

Quarkonium designates a flavorless meson whose constituents are a
quark and its own antiquark i-e $Q\bar{Q},$ $Q=c,b$, is called
quarkonium e.g. charmonium $c\bar{c}$, bottomonium $b\bar{b}$ .
Examples of quarkonia are the $J/\Psi $ and $\Upsilon (nS)$ .
Because of the high mass of the top quark, a toponium does not
exist, since the quark decays through the electroweak interaction
before a bound state can form.

For many reasons the strong interactions of hadrons containing heavy
quarks are easier to understand than those of hadrons containing
only light quarks. The first is asymptotic freedom, the fact that
the effective coupling constant of QCD becomes weak in processes
with large momentum transfer, corresponding to interactions at short
distance scales. At large distances, on the other hand, the coupling
becomes strong, leading to nonperturbative
phenomena such as the confinement of quarks and gluons on a length scale $%
R_{had}\sim 1/\Lambda _{QCD}\sim 1fm$, which determines the size of
hadrons. Roughly speaking, $\Lambda _{QCD}\sim 0.2GeV\ $is the
energy scale that separates the regions of large and small coupling
constant. When the mass of a quark $Q$ is much larger than this
scale, $m_{Q}\gg \Lambda _{QCD}$, it is called a heavy quark.

The light degrees of freedom are blind to the flavor (mass). This is
known as \emph{flavor symmetry}. The heavy quark spin also decouples
from the strong interaction. The decoupling of the spin in the heavy
quark limit leads to the \emph{spin symmetry}. These two symmetries
have important consequences, especially for the decays of beauty
hadrons to lighter hadrons. These symmetries are only true in the
heavy quark limit and are violated at order $\frac{\Lambda
_{QCD}}{m_{Q}}$.

Since quarks are fermions with spin $1/2$, the wave function is
antisymmetric with the exchange of particles $Q$ and $\bar{Q}$.
Under
particle exchange, we get with space coordinates exchange, a factor $%
(-1)^{L} $, with spin coordinates exchange, a factor $(-1)^{S+1}$
and with charge exchange, a factor $C$ ($C$ is called $C$-parity).
Hence Pauli principle gives

\begin{equation}
(-1)^{L+S+1}C=-1  \label{1.13}
\end{equation}%
In the spectroscopic notation, state is completely specified as%
\begin{equation}
n\text{ }^{2S+1}L_{J}  \label{1.14}
\end{equation}%
where $n$ is the principal quantum number and $J$ is the total
angular
momentum. Thus for $L=0$, we have the following states%
\begin{eqnarray*}
n\text{ }^{1}S_{0}\text{ \ \ }C &=&+1,\text{ \ \ }n=1,2,.... \\
n\text{ }^{1}S_{0}\text{ \ \ }C &=&-1,\text{ \ \ }n=1,2,....
\end{eqnarray*}%
The ground state is therefore a hyperfine doublet $1^{1}S_{0}(0^{-+})$ and $%
1^{3}S_{1}(1^{--})$. For $L=1$, we have the following states%
\begin{eqnarray*}
n\text{ }^{1}P_{J}\text{ \ \ }J &=&+1\text{ \ \ \ }C=-1\text{ \ \ } \\
n\text{ }^{3}P_{J}\text{ \ \ }J &=&0,1,2\text{ \ \ \ }C=1\text{ \ \
}
\end{eqnarray*}%
Similarly we can write states for $L=2$.

It is noted that the state $^{3}D_{1}$ has the same quantum number as $%
^{3}S_{1}$, therefore they can mix. Most of these states have been
discovered experimentally \cite{6}. Some of the states are
predicted, but have not been identified; others are unconfirmed.

The computation of the properties of mesons in Quantum
chromodynamics (QCD) is a fully non-perturbative one. As a result,
the only general method available is a direct computation using
Lattice QCD (LQCD) techniques. However, other techniques are
effective for heavy quarkonia as well. The speed of the charm and
the bottom quarks in their respective quarkonia is sufficiently
smaller, so that relativistic effects in these states are much less.
This technique is called non-relativistic QCD (NRQCD).

All known hadrons are color singlets. The exchange of gluons can
provide binding between quarks in a hadron. The two body one gluon
exchange color electric potential is given by:

\begin{eqnarray}
V_{ij} &=&k_{s}\frac{\alpha _{s}}{r}\text{ },k_{s}=\{-\frac{4}{3}\text{ }for%
\text{ }q\bar{q},\text{ }  \label{1.15} \\
V_{ij} &=&k_{s}\frac{\alpha _{s}}{r}\text{ },k_{s}=\{-\frac{2}{3}\text{ }for%
\text{ }qq  \label{1.16}
\end{eqnarray}%
Here $i$, $j$ are flavor indices. Since the running coupling
constant becomes smaller as we decrease the distance, the effective
potential $V_{ij}$ in the lowest order as given by one-gluon
exchange potential is a very good approximation for short distances.
We conclude that for short distances, one can use the one gluon
exchange potential, taking into account the running coupling
constant $\alpha _{s}.$ The second regime, i.e. for large $r,$ QCD
perturbation theory breaks down and we have the confinement of the
quarks. One may look for the origin of this yet unsatisfactorily
explained phenomena. There are many pictures which support the
existence of a linear confining term. One of them is the string
picture of hadrons.

An early, but still effective, technique uses models of the
effective potential to calculate masses of quarkonia states. In this
technique, one uses the fact that the motion of the quarks that
comprise the quarkonium state is nonrelativistic to assume that they
move in a static potential, much like nonrelativistic models of the
hydrogen atom. One of the most
popular potential models is the so-called Cornell potential%
\begin{equation}
V(r)=\frac{a}{r}+br  \label{1.17}
\end{equation}%
where $r$ is the effective radius of the quarkonium state $a$ and
$b$ are parameters. The first part, $a/r$ corresponds to the
potential induced by one-gluon exchange between the quark and its
anti-quark, and is known as the Coulombic part of the potential. The
second part, $br$ the linear term is the phenomenological
implementation of the confining force between quarks, and
parameterizes the poorly-understood non-perturbative effects of QCD.
Relativistic and other effects can be incorporated into this
approach by adding extra terms to the potential.

\chapter{TETRAQUARKS: THE $X\left( 3872\right) $}

\section{Facts About The\textbf{\ }$X\left( 3872\right) $}

The $X$ was found in an exclusive decay in August $2003$ \cite{16}.%
\begin{equation}
B^{+}\rightarrow K^{+}X\left( 3872\right) \rightarrow K^{+}J/\psi
\pi ^{+}\pi ^{-}  \label{2.1}
\end{equation}%
Belle measured its mass:%
\begin{equation}
m(X\left( 3872\right) )=3872.0\pm 0.6(stat.)\pm 0.5\text{
}MeV/c^{2}(syst.) \label{2.2}
\end{equation}%
Belle also set a limit on its decay width:

\begin{equation}
\Gamma (X\left( 3872\right) )<2.3\text{ }MeV\text{ }  \label{2.3}
\end{equation}%
The most natural interpretation was a new coventional$\ c\bar{c}$
state. These exotic states refuse to obey conventional c\={c}
charmonium
interpretation \cite{28,29}. So many theories came: molecular models \cite%
{25,26}, more general 4-quark interpretations including a
diquark-antiquark model \cite{17}-\cite{21}, hybrid models \cite{27}
etc, to explain the nature of the $X\left( 3872\right) .$ In this
chapter we mainly focus on the diquark-antidiquark model.

\subsection{Charmonium Hypothesis}

The fact that the $X\left( 3872\right) $\ decays to $J/\psi \pi
^{+}\pi ^{-}$ suggests that it contains $c$ and $\bar{c}$ quarks,
and the most natural choice to explain the $X\left( 3872\right) $ is
an unseen charmonium state. Charmonium system has been thoroughly
studied in \cite{23}. After the discovery of the $X\left(
3872\right) $ more and more similar narrow resonances have been
discovered and confirmed at electron-positron and proton-antiproton
colliders. Twenty new unexpected charmonium-like particles have been
found which have clear clashes with standard charmonium
interpretations. Let us discuss the conventional charmonium states
and the main objection for assigning them to the $X(3872).$

The lightest charmonium state is called $\eta _{c}.$ This is the
S-state in which the spins of the quarks are antiparallel, so that
the total spin is
zero and the total angular momentum $J=0.$ The radial quantum number of the $%
\eta _{c}$\ is $n=1$, so that the spectroscopic notation ($n$ $%
^{2S+1}L_{J^{PC}}$) for this state is $1^{1}S_{0}$ and the $J^{PC}$ is $%
0^{-+}.$ The next, a bit heavier, state is the $J/\psi .$ The next $c\bar{c}$%
\ state $\chi _{c0}$ is a $P-$wave, more exactly $1^{3}P_{0}$, with $%
J^{PC}=0^{++}.$ This is a part of a triplet, three particles with the same $%
L $ and $S,$ but different $J.$ The other two particles in the triplet are $%
\chi _{c1}(1^{3}P_{1^{++}})$ and $\chi _{c2}(1^{3}P_{2^{++}}).$ We
discuss the upper part of the spectrum. Only a few of the charmonia
states with masses above the $D\bar{D}$ threshold are cosidered.
There are a few interesting states e.g. $^{1}D_{2}$ and $^{3}D_{2}$.
The decay of these
states into $D\bar{D}$ is forbidden because of their spin-parity $%
J^{PC}=2^{-\pm }.$ Both $D$ and $\bar{D}$ have zero spin and the
spin-parity of the $D\bar{D}$\ system is determined by the same way
as for two pions.
The total even $J$ of $D\bar{D}$\ system constrains the $C-$ and $P-$%
parities to be positive. Therefore these states cannot decay into
$D\bar{D}$ and expected to have small widths.

As mentioned earlier that initially the $X\left( 3872\right) $ was
expected to be one of the so far unknown higher mass charmonium
states. However, interpreting the $X\left( 3872\right) $ as a
conventional state is problematic. We go through all the $c\bar{c}$\
states which are not yet identified and evaluate by their
suitability for the $X\left( 3872\right) .$ The states $2S,$ $3P,$
$3D$ and higher are expected to be much heavy to
associated with the $X\left( 3872\right) .$ We do not consider $1S,$ $1P$ and%
$\ 2S,$ because they are unambiguously identified already. Ten
states
remain, two of them are known: 1$^{3}D_{1^{--}}$ is $\psi (3770)$ and $%
3^{3}S_{1^{--}}$ is $\psi (4040)$, so we will not consider them as
serious candidates for the $X\left( 3872\right) .$ Now the eight
states remain. The remaining eight states can not be interpreted as
$X$ \ \cite{5}, as mention in the following Table. To be precise,
$X\left( 3872\right) $ is not a conventional charmonium state.

\begin{table}[h]
\caption{Conventional charmonium states and the main objection for
assigning them to the $X(3872).$}
\begin{center}
\begin{tabular}{||l||l||l||l||}
\hline\hline $1$ & $2$ & $3$ & $4$ \\ \hline\hline
$n^{2S+1}L_{J^{PC}}$ &
\begin{tabular}{l}
Mass \\
$MeV/c^{2}$%
\end{tabular}
&
\begin{tabular}{l}
$\pi ^{-}\pi ^{+}$ \\
$J^{PC}$%
\end{tabular}
& Main objections for the $X\left( 3872\right) $\ asignment \\
\hline\hline
\begin{tabular}{l}
$1^{1}D_{2^{-+}}$ \\
$1^{3}D_{2^{--}}$ \\
$1^{3}D_{3^{--}}$%
\end{tabular}
&
\begin{tabular}{l}
$\sim 3838$ \\
$\sim 3838$ \\
$\sim 3838$%
\end{tabular}
&
\begin{tabular}{l}
$1^{--}$ \\
$0^{++}$ \\
$0^{++}$%
\end{tabular}
&
\begin{tabular}{l}
expect $\eta _{c}\pi \pi \gg J/\psi \pi \pi $ \\
not seen decay to $\chi _{c1}\gamma $ \\
not seen decay to $\chi _{c2}\gamma $%
\end{tabular}
\\ \hline\hline
\begin{tabular}{l}
$2^{1}P_{1^{+-}}$ \\
$2^{3}P_{0^{++}}$ \\
$2^{3}P_{1^{++}}$ \\
$2^{3}P_{2^{++}}$%
\end{tabular}
&
\begin{tabular}{l}
$\sim 3968$ \\
$\sim 3932$ \\
$\sim 4008$ \\
$\sim 3966$%
\end{tabular}
&
\begin{tabular}{l}
0$^{++}$ \\
1$^{--}$ \\
1$^{--}$ \\
1$^{--}$%
\end{tabular}
&
\begin{tabular}{l}
wrong cos$\theta _{J/\psi }$ distribution \\
$D\bar{D}$ not suppressed$\rightarrow $broad \\
too large expected width to $J/\psi \gamma $ \\
$D\bar{D}$ not suppressed$\rightarrow $broad%
\end{tabular}
\\ \hline\hline
$3^{1}S_{0^{-+}}$ & $\sim 4040$ & 1$^{--}$ & mass expected to be close to $%
^{3}S_{1}$ \\ \hline\hline
\end{tabular}%
\end{center}
\end{table}

\subsection{Weakly Bound $D-D^{\ast }$ State}

There are two possibilities to form bound states out of two quarks
and two
anti-quarks: In this configuration, binding each quark to an anti-quark $%
\left[ q_{\alpha }\bar{q}^{\alpha }\right] $ and allowing
interaction
between the two color neutral pairs $\left[ q_{\alpha }\bar{q}^{\alpha }%
\right] \left[ q_{\beta }\bar{q}^{\beta }\right] $. Other possible
configuration will be discussed in the next section. The mainstream
thought has been that of identifying most of these resonances as
molecules of charm mesons. In particular the $X\left( 3872\right) $
happens to have a mass very
close to the $D\bar{D}^{\ast }$ threshold. The binding energy left for the $%
X\left( 3872\right) $ is consistent with $E\thicksim 0.25\pm $
$0.40$ MeV, thus making this state very large in size: order of ten
times bigger than the typical range of strong interactions. These
questions apply to other near-to-threshold hypothetical molecules
and have induced to think to alternative explanations for the
$X\left( 3872\right) $ and its relatives.

\section{Diquark-Antidiquark Model}

In the diquark-antidiquark model the mass spectra are computed as in
the non-relativistic constituent quark model. In the constituent
quark model hadron masses are described by an effective Hamiltonian
that takes as input the constituent quark masses and the spin-spin
couplings between quarks. By extending this approach to
diquark-antidiquark bound states it is possible
to predict tetraquark mass spectra. The mass spectrum of tetraquarks $%
\mathcal{Q\bar{Q}}$ with $\mathcal{Q}=[qq]$ can be described in
terms of the constituent diquark masses, $m_{\mathcal{Q}}$,
spin-spin interactions inside the single diquark, spin-spin
interaction between quark and antiquark belonging to two diquarks,
spin-orbit, and purely orbital term \cite{17}. This model has been
tested on standard charmonia and has a rather good behavior to
determine the mass spectra.

\section{Constituent Quarks and Spin-Spin Interactions}

In the costituent quark model the Hamiltonian is \cite{59}:

\begin{equation}
H=\sum\limits_{i}m_{i}+\sum\limits_{i<j}2\mathcal{K}_{ij}(\mathbf{S}%
_{i}\cdot \mathbf{S}_{j})  \label{2.4}
\end{equation}%
where the sum runs over the hadron constituents. The coefficient $\mathcal{K}%
_{ij}$ depends on the flavor of the constituents $i$, $j$ and on the
particular color state of the pair. Couplings for color singlet
combinations
are determined from the scalar and vector light mesons. For the $L=0\ $%
mesons, taking $u\bar{s}$ states, Eq.(\ref{2.4}) gives

\begin{equation}
M=m_{q}+m_{s}+\mathcal{K}_{s\bar{q}}[J(J+1)-\frac{3}{2}]
\label{2.5}
\end{equation}%
\[
M_{K}=m_{q}+m_{s}+\mathcal{K}_{s\bar{q}}[-\frac{3}{2}]
\]%
Similarly for the vector meson $K^{\ast }$%
\[
M_{K^{\ast }}=m_{q}+m_{s}+\mathcal{K}_{s\bar{q}}[1(1+1)-\frac{3}{2}]
\]

\[
M_{K^{\ast }}=m_{q}+m_{s}+\mathcal{K}_{s\bar{q}}[\frac{1}{2}]
\]

\[
M_{K^{\ast }}-M_{K}=2\mathcal{K}_{s\bar{q}},\
\mathcal{K}_{s\bar{q}}=195\ MeV
\]%
Adding the similar equations for $\pi -\rho $, $D-D^{\ast }$, $%
D_{s}-D_{s}^{\ast }$, $J/\psi -\eta _{c}$ complex we obtain the
values of the spin-spin couplings, for quark-antiquark pairs in the
color singlet state from known $L=0$\ mesons.
\begin{table}[tb]
\caption{Constituent quark masses derived from the $L=0$ mesons and
baryons.} \label{Table I}
\begin{center}
\begin{tabular}{|l|l|l|l|l|}
\hline Constituent mass (MeV) & $q$ & $s$ & $c$ & $b$ \\ \hline
Mesons & $305$ & $490$ & $1670$ & $5008$ \\ \hline Baryons & $362$ &
$546$ & $1721$ & $5050$ \\ \hline\hline
\end{tabular}%
\end{center}
\end{table}

\begin{table}[tb]
\caption{Spin-Spin couplings for quark-antiquark pairs in in the
color singlet state from the known\ mesons.}
\begin{center}
$%
\begin{tabular}{||l||l||l||l||l||l||l||}
\hline\hline
Spin-spin couplings & q\={q} & s\={q} & s\={s} & c\={q} & c\={s} & c\={c} \\
\hline\hline $(K_{ij})_{0}$(MeV) & 318 & 200 & 129 & 71 & 72 & 59 \\
\hline\hline
\end{tabular}%
$%
\end{center}
\end{table}
Now spin-spin coupling for quark-quark in color $\bar{3}$
(antitriplet) state can be calculated from the known $L=0$ baryons.
The $qq$ couplings are determined from the masses of the $qqq$
baryons ground $(J=1/2)$ and excited
$(J=3/2)$ states. Lets take the \textit{uds }states:$\Lambda $, $\Sigma $, $%
\Sigma ^{\ast },$ which gives%
\begin{equation}
M=2m_{q}+m_{s}+(\mathcal{K}_{qq})_{_{\bar{3}}}[S(S+1)-\frac{3}{2}]+(\mathcal{%
K}_{qs})_{_{\bar{3}}}[J(J+1)-S(S+1)-\frac{3}{4}]  \label{2.6}
\end{equation}%
Writing equations for $P,$ $\Delta ^{+}$, involving only$\ (\mathcal{K}%
_{qq})_{_{\bar{3}}}$, and for $\Lambda _{c},\Sigma _{c},\Sigma
_{c}^{\ast }$
, involving $(\mathcal{K}_{qq})_{_{\bar{3}}}$ and $(\mathcal{K}_{qc})_{_{%
\bar{3}}}$. Also we consider the three $\Xi _{c}\ $states which\ give $(%
\mathcal{K}_{qc})_{_{\bar{3}}}$ and
$(\mathcal{K}_{sc})_{_{\bar{3}}}$ couplings. These are given in
Table 3.4 where one can see that the coupling strength decreases
with increasing mass.

\begin{table}[tb]
\caption{Spin-Spin couplings for quark-quark in color $\bar{3}$
state from the known baryons.}
\begin{center}
$%
\begin{tabular}{||l||l||l||l||l||}
\hline\hline Spin-spin couplings & $qq$ & $sq$ & $cq$ & $cs$ \\
\hline\hline $(K_{ij})_{\bar{3}}$(MeV) & $98$ & $65$ & $22$ & $24$
\\ \hline\hline
\end{tabular}%
$%
\end{center}
\end{table}
It is observed that the diquark correlation decreases when one of
the light quarks is strange. According to one gluon exchange (c.f \
Eq.(\ref{1.15}) and Eq.(\ref{1.16})), we have

\begin{equation}
(\mathcal{K}_{ij})_{\bar{3}}=\frac{1}{2}(\mathcal{K}_{ij})_{0}
\label{2.7}
\end{equation}%
This relates coupling of antitriplet to the singlet state.

The couplings corresponding to the spin-spin interactions have been
calculated for the color singlet and color antitriplet only. The
couplings are not necessarily in the singlet state but octet
couplings $\left(
\mathcal{K}_{c\bar{c}}\right) _{8}$ are also possible. The quantities $%
\mathcal{K}_{q\bar{q}}$, $\mathcal{K}_{c\bar{q}}$ and $\mathcal{K}_{c\bar{c}%
} $ involve both color singlet and color octet couplings between the
quarks and antiquraks in a $\mathcal{Q\bar{Q}}$ system (A quark in
the\ diquark $\mathcal{Q}$ could have a color octet spin-spin
interaction with an antiquark in the antidiquark
$\mathcal{\bar{Q}}$). For the diquark
attraction in the $\bar{3}-$color state, we can write $\mathcal{Q}%
^{i}=[cq]^{i}=\epsilon ^{ijk}c_{j}q_{k}$, where $i,\ j,\ k\ $are
color indices in the fundamental representation of $SU(3).$ The
color singlet hadron is written as
\begin{equation}
\lbrack cq][\bar{c}\bar{q}]=\epsilon ^{ijk}\epsilon _{ij^{\prime
}k^{\prime
}}(c_{j}q_{k})(\bar{c}^{j^{\prime }}\bar{q}^{k^{\prime }})=(c_{j}\bar{c}%
^{j})(q_{k}\bar{q}^{k})-(c_{j}\bar{q}^{j})(q_{k}\bar{c}^{k})
\label{2.8}
\end{equation}%
Rearranging the color indices in the last term by using $SU(N)$
identity for
the Lie algebra generators%
\begin{equation}
\sum\limits_{a=1}^{N^{2}-1}\lambda _{ij}^{a}\lambda
_{kl}^{a}=2(\delta _{il}\delta _{jk}-\frac{1}{N}\delta _{ij}\delta
_{kl})  \label{2.9}
\end{equation}%
where $N$\ is the number of colors. A color octet $(N\ =3)$
$q\bar{q}$ state
can be written as $\bar{q}^{i}\lambda _{ij}^{a}q^{j},$ and hence%
\begin{equation}
(\bar{c}^{i}\lambda _{ij}^{a}c^{j})(\bar{q}^{k}\lambda
_{kl}^{a}q^{l})=\sum\limits_{a=1}^{N^{2}-1}\lambda _{ij}^{a}\lambda
_{kl}^{a}(c_{j}\bar{c}^{j})(q_{k}\bar{q}^{k})=2\left[ (c_{j}q_{k})(\bar{c}%
^{j^{\prime }}\bar{q}^{k^{\prime }})-\frac{1}{N}(c_{j}\bar{c}^{j})(q_{k}\bar{%
q}^{k})\right]  \label{2.10}
\end{equation}%
Using Eq.(\ref{2.9}) and Eq.(\ref{2.10}), we extract the octet term
as
follows:%
\begin{eqnarray}
\lbrack cq][\bar{c}\bar{q}]
&=&(c_{j}\bar{c}^{j})(q_{k}\bar{q}^{k})-\left[
\frac{1}{2}(\bar{c}^{i}\lambda _{ij}^{a}c^{j})(\bar{q}^{k}\lambda
_{kl}^{a}q^{l})+\frac{1}{3}(c_{j}\bar{c}^{j})(q_{k}\bar{q}^{k})\right]
\label{2.11} \\
&=&\frac{2}{3}(c_{j}\bar{c}^{j})(q_{k}\bar{q}^{k})-\frac{1}{2}(\bar{c}%
^{i}\lambda _{ij}^{a}c^{j})(\bar{q}^{k}\lambda _{kl}^{a}q^{l})
\label{2.12}
\end{eqnarray}%
This formula gives information about the relative weights of a
singlet and an octet color state in a diquark-antidiquark picture.
We have three colors
running in the sum $c_{i}\bar{c}^{i}$ whereas $a=1,...,8$ in $\bar{c}%
^{i}\lambda _{ij}^{a}c^{j}$. Therefore the probability of finding a
particular $q\bar{q}$\ pair in color singlet, for example $c\bar{c}$
in the color singlet state $c_{j}\bar{c}^{j}$, is half the
probability of finding the same pair in color octet $\bar{c}\lambda
^{a}c$ i-e, $3\times \ 2/3=1/2(8\times \ 1/2).$

We write for $\mathcal{K}_{c\bar{c}}$ \cite{17}:
\begin{equation}
\mathcal{K}_{c\bar{c}}\left( [cq][\bar{c}\bar{q}]\right)
=\frac{1}{3}\left(
\mathcal{K}_{c\bar{c}}\right) _{0}+\frac{2}{3}\left( \mathcal{K}_{c\bar{c}%
}\right) _{8}  \label{2.13}
\end{equation}%
where $\left( \mathcal{K}_{c\bar{c}}\right) _{0}$ is reported in Table 3.3. $%
\left( \mathcal{K}_{c\bar{c}}\right) _{8}$ can be derived from the
one gluon
exchange model by using the relation \cite{18}:%
\begin{equation}
\left( \mathcal{K}_{c\bar{c}}\right) _{\mathbf{X}}\sim \left(
C^{2}\left(
\mathbf{X}\right) -C^{2}\left( \mathbf{3}\right) -C^{2}\left( \mathbf{\bar{3}%
}\right) \right)  \label{2.14}
\end{equation}%
where $\mathbf{X}$ is the color representation of the two quark
system, with
$C^{2}\left( \mathbf{X}\right) =0$, $4/3$, $4/3$, $3$ for $\mathbf{X=0}$, $%
\mathbf{3}$, $\mathbf{\bar{3}}$, $\mathbf{8}$ respectively. It is
found that

\begin{equation}
(\mathcal{K}_{c\bar{c}})_{0}\sim -\frac{8}{3}\ \ \ \ \ \ (\mathcal{K}_{c\bar{%
c}})_{8}\sim \frac{1}{3}=-\frac{1}{8}(\mathcal{K}_{c\bar{c}})_{0}
\label{2.15}
\end{equation}%
Finally, from\ Eq.(\ref{2.13}), one has

\[
\mathcal{K}_{c\bar{c}}\left( [cq][\bar{c}\bar{q}]\right)
=\frac{1}{4}\left( \mathcal{K}_{c\bar{c}}\right) _{0}
\]%
Now we have all the couplings and let us apply it to calculate the
mass of light diquark for a simple case of $a_{0}(980)$:

\[
a_{0}(980)=[sq]_{S=0}[\bar{s}\bar{q}]_{S=0}
\]%
Using the Eq.(\ref{2.5}) and eigen state given in the above
equation, we calculate the mean value

\begin{eqnarray*}
\langle a_{0}\left\vert H\right\vert a_{0}\rangle &=&2m_{[sq]}+(\mathcal{K}%
_{sq})_{\bar{3}}[-\frac{3}{2}-\frac{3}{2}]=2m_{[sq]}-3(\mathcal{K}_{sq})_{%
\bar{3}} \\
m_{[sq]} &=&595\ \ MeV
\end{eqnarray*}%
Similarly one can calculate the $m_{[ud]}=396\ \ MeV$ from $\sigma
(481)$.

\section{Spectrum Of Hidden\ Charm Diquark-antidiquark States}

In the diquark-antidiquark model effective Hamiltonian takes the
form:
\begin{equation}
H=2m_{\mathcal{Q}}+H_{SS}^{(\mathcal{QQ)}}+H_{SS}^{(\mathcal{Q\bar{Q}%
\mathcal{)}}}+H_{SL}+H_{LL}  \label{2.16}
\end{equation}%
where $m_{\mathcal{Q}}$ is the mass of diquark,
$H_{SS}^{(\mathcal{QQ)}}$ is
the spin-spin interaction inside the single diquark, $H_{SS}^{(\mathcal{Q%
\bar{Q}\mathcal{)}}}$ is the spin-spin interaction between quark and
antiquark belonging to two diquarks, $H_{SL}$ is the spin-orbit, and
$H_{LL}$
is purely orbital term \cite{17} i.e%
\begin{eqnarray}
H_{SS}^{(\mathcal{QQ)}} &=&2(\mathcal{K}_{cq})_{\bar{3}}[(\mathbf{S}%
_{c}\cdot \mathbf{S}_{q})+(\mathbf{S}_{\bar{c}}\cdot
\mathbf{S}_{\bar{q}})],
\label{2.16(a)} \\
H_{SS}^{(\mathcal{Q\bar{Q}\mathcal{)}}} &=&2(\mathcal{K}_{c\bar{q}})(\mathbf{%
S}_{c}\cdot \mathbf{S}_{\bar{q}}+\mathbf{S}_{\bar{c}}\cdot
\mathbf{S}_{q})+
\nonumber \\
&&2\mathcal{K}_{c\bar{c}}(\mathbf{S}_{c}\cdot \mathbf{S}_{\bar{c}})+2%
\mathcal{K}_{q\bar{q}}(\mathbf{S}_{q}\cdot \mathbf{S}_{\bar{q}}),
\label{(2.16(b))} \\
H_{SL} &=&2A_{\mathcal{Q}}(\mathbf{S}_{\mathcal{Q}}\cdot \mathbf{L}+\mathbf{S%
}_{\mathcal{\bar{Q}}}\cdot \mathbf{L}),  \label{2.16(c)} \\
H_{LL} &=&B_{\mathcal{Q}}\frac{L(L+1)}{2}.  \label{2.16(d)}
\end{eqnarray}%
The overall factor of $2$ is just a convention used in the literature. $A_{%
\mathcal{Q}},\ B_{\mathcal{Q}}\ $are coefficients to be calculated
by using known data. We will use these values in the next chapter.

To calculate the spin-spin interaction of the $\mathcal{Q\bar{Q}}$
states explicitly, we use the following non-relativistic notation
for labelling the state
\begin{equation}
\left\vert
S_{\mathcal{Q}}\text{,~}S_{\bar{\mathcal{Q}}};~J\right\rangle
=\left\vert \Gamma \text{,~}\Gamma ^{\prime };~J\right\rangle
=(c^{a}\Gamma _{ab}u^{b})(\bar{c}^{c}\Gamma _{cd}^{\prime
}\bar{u}^{d})  \label{2.17}
\end{equation}%
where, $S_{\mathcal{Q}}$ and $S_{\bar{\mathcal{Q}}}$ are the spin of
diquark
and antidiquark, respectively, $J$ is the total angular momentum and the $%
\Gamma ^{\alpha }$ are $2\times 2$ matrices in spinor space. Using
Pauli matrices these can be written as:
\begin{equation}
\Gamma ^{0}=\frac{\sigma _{2}}{\sqrt{2}};~\Gamma ^{i}=\frac{1}{\sqrt{2}}%
\sigma _{2}\sigma _{i}~  \label{2.18}
\end{equation}%
for spin 0 and 1, respectively. The matrices $\Gamma $ are
normalised so
that:%
\[
Tr[(\Gamma ^{\alpha })^{\dagger }(\Gamma ^{\beta })]=\delta ^{\alpha
\beta }
\]%
We define the spinor operators as:%
\[
\mathbf{S}_{u}\left\vert \Gamma \right\rangle \equiv \left\vert \Gamma \frac{%
1}{2}\sigma \right\rangle ;\ \ \ \ \mathbf{S}_{c}\left\vert \Gamma
\right\rangle \equiv \left\vert \frac{1}{2}\sigma ^{T}\Gamma
\right\rangle
\]%
since,%
\[
\sigma ^{T}\sigma _{2}=-\sigma _{2}\sigma
\]%
We calculated formula for total spin operator as expected:%
\begin{eqnarray}
(\mathbf{S}_{u}\mathbf{+\ S}_{c})\left\vert \Gamma ^{0}\right\rangle
&=&0
\label{2.19} \\
\lbrack \mathbf{(S}_{u}\mathbf{)}^{i}\mathbf{+(\ S}_{c}\mathbf{)}%
^{i}]\left\vert \Gamma ^{j}\right\rangle &=&i\epsilon
^{ijk}\left\vert \Gamma ^{k}\right\rangle  \label{2.20}
\end{eqnarray}%
We also find:%
\begin{eqnarray}
\langle 0|\ \mathbf{S}_{u}\left\vert 1\right\rangle &=&-\langle 0|\ \mathbf{S%
}_{c}\left\vert 1\right\rangle =\frac{1}{2};  \label{2.21} \\
\langle 1|\ \mathbf{S}_{u}\left\vert 1\right\rangle &=&\langle 1|\ \mathbf{S}%
_{c}\left\vert 1\right\rangle =\frac{1}{2}\langle 1|(\mathbf{S}_{u}\mathbf{+S%
}_{c})\left\vert 1\right\rangle  \label{2.22}
\end{eqnarray}%
We have used the following Pauli matrices properties:%
\begin{eqnarray*}
\lbrack \sigma ^{i},\sigma ^{j}] &=&2i\epsilon ^{ijk}\sigma ^{k} \\
Tr(\sigma ^{i}\sigma ^{i}) &=&Tr(I)=2
\end{eqnarray*}%
By using this information we now calculate the matrix elements of
products of spin operators. There are two cases.

\textbf{Same diquark, }e.g\textbf{.}$S_{u}\cdot S_{c}.$

This operator is only a combination of Casimir operators and is
diagonal in
the basis.%
\begin{equation}
2(\mathbf{S}_{u}\mathbf{\cdot S}_{c})=(\mathbf{S}_{cu}\mathbf{)}^{2}\mathbf{%
-(S}_{c}\mathbf{)}^{2}\mathbf{-(S}_{u}\mathbf{)}^{2}  \label{2.23}
\end{equation}

\textbf{Different diquarks, }e.g\textbf{.}$S_{u}\cdot \
S_{\bar{u}}.$First
we consider$\ J=0$ states, represented by%
\begin{eqnarray*}
\left\vert 0_{\mathcal{Q}},0_{\bar{\mathcal{Q}}};~0_{J}\right\rangle &=&%
\frac{1}{2}\left( \sigma _{2}\right) \otimes \left( \sigma _{2}\right) , \\
\left\vert 1_{\mathcal{Q}},1_{\bar{\mathcal{Q}}};~0_{J}\right\rangle &=&%
\frac{1}{2\sqrt{3}}\left( \sigma _{2}\sigma ^{i}\right) \otimes
\left( \sigma _{2}\sigma ^{i}\right) ,
\end{eqnarray*}%
Using the basic definitions, we get:%
\begin{eqnarray}
2(\mathbf{S}_{u}\cdot \mathbf{\ S}_{\bar{u}})\left\vert 0_{\mathcal{Q}},0_{%
\bar{\mathcal{Q}}};~0_{J}\right\rangle &=&\frac{1}{4}\left( \sigma
_{2}\sigma ^{i}\right) \otimes \left( \sigma _{2}\sigma ^{i}\right) =\frac{%
\sqrt{3}}{2}\left\vert 1_{\mathcal{Q}},1_{\bar{\mathcal{Q}}%
};~0_{J}\right\rangle  \label{2.24} \\
2(\mathbf{S}_{u}\cdot \mathbf{\ S}_{\bar{u}})\left\vert 1_{\mathcal{Q}},1_{%
\bar{\mathcal{Q}}};~0_{J}\right\rangle &=&\frac{1}{4\sqrt{3}}\left(
\sigma _{2}\sigma ^{i}\sigma ^{j}\right) \otimes \left( \sigma
_{2}\sigma
^{i}\sigma ^{j}\right)  \nonumber  \label{2.25} \\
&=&\frac{\sqrt{3}}{2}\left\vert 0_{\mathcal{Q}},0_{\bar{\mathcal{Q}}%
};~0_{J}\right\rangle -\left\vert 1_{\mathcal{Q}},1_{\bar{\mathcal{Q}}%
};~0_{J}\right\rangle
\end{eqnarray}
which leads to the following matrices:%
\begin{equation}
2(S_{u}\cdot \ S_{\bar{u}})=\left(
\begin{array}{cc}
0 & \frac{\sqrt{3}}{2} \\
\frac{\sqrt{3}}{2} & -1%
\end{array}%
\right)  \label{2.26}
\end{equation}%
Now we consider$\ J=1$ states, given in the tensor basis:
\begin{eqnarray}
\left\vert 0_{\mathcal{Q}},1_{\bar{\mathcal{Q}}};~1_{J}\right\rangle &=&%
\frac{1}{2}\left( \sigma _{2}\right) \otimes \left( \sigma
_{2}\sigma
^{i}\right) ,  \nonumber \\
\left\vert 1_{\mathcal{Q}},0_{\bar{\mathcal{Q}}};~1_{J}\right\rangle &=&%
\frac{1}{2}\left( \sigma _{2}\sigma ^{i}\right) \otimes \left(
\sigma
_{2}\right) ,  \nonumber \\
\left\vert 1_{\mathcal{Q}},1_{\bar{\mathcal{Q}}};~1_{J}\right\rangle &=&%
\frac{1}{2\sqrt{2}}\varepsilon ^{ijk}\left( \sigma _{2}\sigma
^{j}\right) \otimes \left( \sigma _{2}\sigma ^{k}\right) .
\end{eqnarray}%
The normalisation of the Hamiltonian in Eq.(\ref{2.16}), using the
basis of states defined above with definite diquark and antidiquark
spin and total angular momentum, will give the spectrum of
diquark-antiquark states. There
are two different possibilities: Lowest lying $[cq][\bar{c}\bar{q}]$ states $%
\left( L_{\mathcal{Q\bar{Q}}}=0\right) $ and higher mass $[cq][\bar{c}\bar{q}%
]$ states $\left( L_{Q\bar{Q}}=1\right) $.

\subsection{Lowest Lying $[cq][\bar{c}\bar{q}]$ States}

In the ground state the two diquarks interact only by spin couplings
because the angular momentum is zero $(L_{\mathcal{Q\bar{Q}}}=0)$.
An effective non-relativistic Hamiltonian can be written including
spin-spin interactions within a diquark and between quarks in
different diquarks. The states can be classified in terms of the
diquark and antidiquark spin, $S_{\mathcal{Q}}$ and
$S_{\bar{\mathcal{Q}}}$, total angular momentum $J$, parity, $P$ and
charge conjugation, $C$. Considering both \textit{good}
$(S_{\mathcal{Q}}=0)$
and \textit{bad }$(S_{\mathcal{Q}}=1)$ diquraks and having $L_{\mathcal{Q%
\bar{Q}}}=0$ we have six possible states which are listed below.

\textbf{i. Two states with }$J^{PC}=0^{++}$\textbf{:}%
\begin{eqnarray}
\left\vert 0^{++}\right\rangle &=&\left\vert 0_{\mathcal{Q}},0_{\bar{%
\mathcal{Q}}};~0_{J}\right\rangle ;  \label{2.27} \\
\left\vert 0^{++\prime }\right\rangle &=&\left\vert 1_{\mathcal{Q}},1_{\bar{%
\mathcal{Q}}};~0_{J}\right\rangle .  \label{2.28}
\end{eqnarray}

\textbf{ii. Three states with }$J=1$\textbf{:}%
\begin{eqnarray}
\left\vert 1^{++}\right\rangle &=&\frac{1}{\sqrt{2}}\left( \left\vert 0_{%
\mathcal{Q}},1_{\bar{\mathcal{Q}}};~1_{J}\right\rangle +\left\vert 1_{%
\mathcal{Q}},0_{\bar{\mathcal{Q}}};~1_{J}\right\rangle \right) ;
\label{2.29} \\
\left\vert 1^{+-}\right\rangle &=&\frac{1}{\sqrt{2}}\left( \left\vert 0_{%
\mathcal{Q}},1_{\bar{\mathcal{Q}}};~1_{J}\right\rangle -\left\vert 1_{%
\mathcal{Q}},0_{\bar{\mathcal{Q}}};~1_{J}\right\rangle \right) ;
\label{2.30} \\
\left\vert 1^{+-\prime }\right\rangle &=&\left\vert 1_{\mathcal{Q}},1_{\bar{%
\mathcal{Q}}};~1_{J}\right\rangle .  \label{2.31}
\end{eqnarray}%
All these states have positive parity as both the \textit{good} and \textit{%
bad} diquarks have positive parity and $L_{\mathcal{Q\bar{Q}}}=0$.
The
difference is in the charge conjugation quantum number, the state $%
\left\vert 1^{++}\right\rangle $ is even under charge conjugation, whereas $%
\left\vert 1^{+-}\right\rangle $ and $\left\vert 1^{+-\prime
}\right\rangle $ are odd.

\textbf{iii. One state with }$J^{PC}=2^{++}$\textbf{:}%
\begin{equation}
\left\vert 2^{++}\right\rangle =\left\vert 1_{\mathcal{Q}},1_{\bar{\mathcal{Q%
}}};~2_{J}\right\rangle .  \label{2.32}
\end{equation}%
Keeping in mind that for $L_{\mathcal{Q\bar{Q}}}=0$ there is no
spin-orbit and purely orbital term, the Hamiltonian of
Eq.(\ref{2.16}) takes the form

\begin{equation}
H=2m_{\mathcal{Q}}+H_{SS}^{(\mathcal{QQ)}}+H_{SS}^{(\mathcal{Q\bar{Q}%
\mathcal{)}}}  \label{2.33}
\end{equation}%
Thus,%
\begin{eqnarray}
H &=&2m_{[qc]}+2(\mathcal{K}_{qc})_{\bar{3}}[(\mathbf{S}_{c}\cdot \mathbf{S}%
_{q})+(\mathbf{S}_{\bar{c}}\cdot \mathbf{S}_{\bar{q}})]+2\mathcal{K}_{q\bar{q%
}}(\mathbf{S}_{q}\cdot \mathbf{S}_{\bar{q}})  \nonumber \\
&&+2(\mathcal{K}_{c\bar{q}})(\mathbf{S}_{c}\cdot \mathbf{S}_{\bar{q}}+%
\mathbf{S}_{\bar{c}}\cdot \mathbf{S}_{q})+2\mathcal{K}_{c\bar{c}}(\mathbf{S}%
_{c}\cdot \mathbf{S}_{\bar{c}}).  \label{2.34}
\end{eqnarray}

The diagonalisation of the this Hamiltonian with the states defined
above gives the eigenvalues which are needed to estimate the masses
of these states. It is calculated that for the $1^{++}$ and $2^{++}$
states the Hamiltonian is diagonal with the eigenvalues
\begin{eqnarray}
M\left( 1^{++}\right) &=&2m_{[cq]}-(\mathcal{K}_{cq})_{\bar{3}}+\frac{1}{2}%
\mathcal{K}_{q\bar{q}}-\mathcal{K}_{c\bar{q}}+\frac{1}{2}\mathcal{K}_{c\bar{c%
}},  \label{2.35} \\
M\left( 2^{++}\right) &=&2m_{[cq]}+(\mathcal{K}_{cq})_{\bar{3}}+\frac{1}{2}%
\mathcal{K}_{q\bar{q}}+\mathcal{K}_{c\bar{q}}+\frac{1}{2}\mathcal{K}_{c\bar{c%
}}.  \label{2.36}
\end{eqnarray}

All other quantities are now specified except the mass of the
constituent diquark. The $X(3872)$ is a
$[cq]_{S=1}[\bar{c}\bar{q}]_{S=0}$ tetraquark. By diagonalizing the
Hamiltonian in Eq.(\ref{2.34}) and, using the spin couplings derived
above, the mass of the diquark $[cq]$ was fixed by using the mass of
$X(3872)$ as input, yielding $m_{[cq]}=1.933$ $GeV$. The $1^{++}$
state is a good candidate\ to explain the properties of $X(3872).$
In order to reduce the experimental information needed we estimate
the remaining diquark masses by substituting the costituent quark
forming the diquark. We have
\begin{eqnarray}
m_{[cs]} &=&m_{[cq]}-m_{q}+m_{s}  \label{2.37} \\
m_{[bq]} &=&m_{[cq]}-m_{c}+m_{b}  \label{2.38} \\
m_{[bs]} &=&m_{[bq]}-m_{q}+m_{s}  \label{2.39}
\end{eqnarray}

\begin{table}[tb]
\caption{Diquark masses}
\begin{center}
$%
\begin{tabular}{||l||l||l||l||l||l||}
\hline\hline Diquark masses & $[ud]$ & $[sq]$ & $[cq]$ & $[cs]$ &
$[bq]$ \\ \hline\hline $MeV$ & $396$ & $595$ & $1933$ & $2118$ &
$5119$ \\ \hline\hline
\end{tabular}%
$%
\end{center}
\end{table}

Now, we have all the input parameters to calculate the mass spectrum
numerically. Putting masses of diquarks from Table (2.4) and values
of couplings from Tables (2.2), (2.3) in Eq.(\ref{2.35}), we get the
mass for the hidden $c\bar{c}$ tetraquark $1^{++}$ state:

\begin{equation}
M\left( 1^{++}\right) =3.872\text{ }GeV  \label{2.40}
\end{equation}%
Taking the\textbf{\ }$X\left( 3872\right) $ as input we can also
predict the
existence of a $2^{++}$ state that can be associated to the \textbf{\ }$%
X\left( 3940\right) $ observed by Belle \cite{51}:
\[
M\left( 2^{++}\right) =3.952\text{ }GeV.
\]%
For the corresponding $0^{++}$ and $1^{+-}$ (Labelled as $Z$ )
tetraquark
states, the Hamiltonian is not diagonal and we calculated the following $%
2\times 2$ matrices, using the non-relativistic notation for
labelling the
state as in Eq.(\ref{2.26}):%
\[
M\left( 0^{++}\right) =\left(
\begin{array}{cc}
-3(\mathcal{K}_{cq})_{\bar{3}} & \frac{\sqrt{3}}{2}\left( \mathcal{K}_{q\bar{%
q}}+\mathcal{K}_{c\bar{c}}-2\mathcal{K}_{c\bar{q}}\right) \\
\frac{\sqrt{3}}{2}\left( \mathcal{K}_{q\bar{q}}+\mathcal{K}_{c\bar{c}}-2%
\mathcal{K}_{c\bar{q}}\right) & (\mathcal{K}_{cq})_{\bar{3}}-\left( \mathcal{%
K}_{q\bar{q}}+\mathcal{K}_{c\bar{c}}+2\mathcal{K}_{c\bar{q}}\right)%
\end{array}%
\right) ,
\]%
\[
M\left( 1^{+-}\right) =\left(
\begin{array}{cc}
-(\mathcal{K}_{cq})_{\bar{3}}+\mathcal{K}_{c\bar{q}}-\frac{\left( \mathcal{K}%
_{q\bar{q}}+\mathcal{K}_{c\bar{c}}\right) }{2} & \mathcal{K}_{q\bar{q}}-%
\mathcal{K}_{c\bar{c}} \\
\mathcal{K}_{q\bar{q}}-\mathcal{K}_{c\bar{c}} & (\mathcal{K}_{cq})_{\bar{3}}-%
\mathcal{K}_{c\bar{q}}-\frac{\left( \mathcal{K}_{q\bar{q}}+\mathcal{K}_{c%
\bar{c}}\right) }{2}%
\end{array}%
\right) .
\]%
To estimate the masses of these two states, one has to diagonalise
the above matrices. After doing this, the mass spectrum of these
states is shown in Fig 3-1.

\begin{figure}[here]
\centering
    \includegraphics[scale=.5]{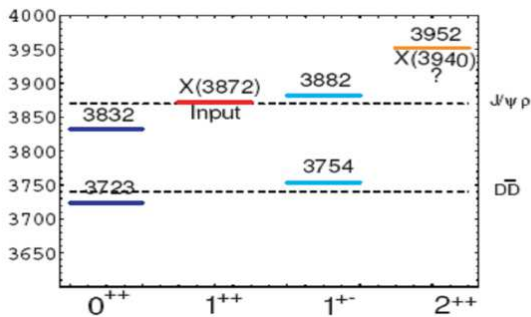}
    \caption{Lowest lying hidden charm
spectrum.}
\end{figure}

\section{Isospin Breaking and Decay Widths of the $J^{PC}=1^{++}$ Tetraquark}

\subsection{Isospin Breaking}

In this section we discuss the isospin breaking effects which were
neglected in the previous section. The isospin quantum number is
related to the finer structure of the $X$ state. The two \ flavor
eigenstates $X_{u}$ and $X_{d}$ mix through self energy diagrams,
which annihilate a $u\bar{u}$ pair and
convert it into a $d\bar{d}$ pair through intermediate gluons. In the basis $%
\{X_{u},X_{d}\}$ the annihilation diagrams contribute equally to all
the entries of the mass matrix, while the contribution of the quark
masses is diagonal. The resulting $2\times 2$ mixing mass\ matrix
is:
\begin{equation}
\left(
\begin{array}{cc}
2m_{u}+\delta & \delta \\
\delta & 2m_{d}+\delta%
\end{array}%
\right)  \label{2.41}
\end{equation}%
where $\delta $ is the contribution from quark annihilation
diagrams. At the scale determined by the $c\bar{c}$ pair the
annihilation term $\delta $ is expected to be small and thus the
mass eigenstates should coincide with \ flavor eigenstates to a
rather good extent. Isospin-breaking introduces a mass splitting and
the mass eigenstates called $X_{[c,l]}$ and $X_{[c,h]}$ (for lighter
and heavier of the two) become linear combinations of $X_{[cu]}$ and
$X_{[cd]}$. One can put:

\begin{eqnarray}
X_{[c,l]} &=&\cos \theta \ X_{[cu]}\;+\sin \theta \;X_{[cd]}  \label{2.42} \\
X_{[c,h]} &=&-\sin \theta \ X_{[cu]}+\cos \theta \ X_{[cd]}
\label{2.43}
\end{eqnarray}%
The mass differences are estimated to be small, where $\theta $ is a
mixing
angle. The electromagnetic couplings of the tetraquarks $X_{[c,l]}$ and $%
X_{[c,h]}$ will depend on the mixing angle $\theta .$

\[
\left(
\begin{array}{c}
X_{[c,l]} \\
X_{[c,h]}%
\end{array}%
\right) =R\left(
\begin{array}{c}
X_{[cu]} \\
X_{[cd]}%
\end{array}%
\right) ,\text{ }R=\left(
\begin{array}{cc}
\cos \theta & \sin \theta \\
-\sin \theta & \cos \theta%
\end{array}%
\right)
\]%
We get,%
\begin{eqnarray*}
2m_{u} &=&M(X_{[c,l]})\cos ^{2}\theta +M(X_{[c,h]})\sin ^{2}\theta \\
2m_{d} &=&M(X_{[c,l]})\sin ^{2}\theta +M(X_{[c,h]})\cos ^{2}\theta
\end{eqnarray*}%
Take the difference:%
\begin{equation}
M(X_{[c,h]})-M(X_{[c,l]})=2(m_{d}-m_{u})/\cos (2\theta )
\label{2.44}
\end{equation}%
Infact, there are two different states $X(3872)$ and $X(3875)$ which
were not excluded from the experimental data \cite{52, 53}. The
isospin violation in the tetraquark picture is the possibility of
$\omega -\rho ^{0}$ mixing,
as proposed in \cite{20}. The observation in $2006$ of a state decaying to $%
D^{0}\bar{D}^{0}\pi $ with mass $3875$ $MeV$ favored the assignment:
$X_{u}$
$=X(3875)$, decaying mainly into $J/\psi \pi ^{+}\pi ^{-}$ and $%
X_{d}=X(3872) $ decaying into $D^{0}\bar{D}^{0}\pi $. The mass
ordering of these two neutral states seems to be reversed, since the
$u$ quark is lighter than the $d$ quark and thus one would expect
$X_{u}$ to be lighter than $X_{d}$. However the quarks which form
the diquarks in the $X_{u}$ have the same electric charge and thus a
consistent consideration of the electrostatic energy can perhaps
change the order of the masses. Besides these two neutral states,
two charged states arise as a natural prediction
of the tetraquark picture $X^{+}=[cu][\bar{c}\bar{d}]$ $and$ $X^{-}=[cd][%
\bar{c}\bar{u}]$. The charged partners are $X^{\pm }$ are not observed \cite%
{30}.

\subsection{Decay Widths of the $J^{PC}=1^{++}$}

Originally the $X(3872)$ was found through its decay into $J/\psi
\pi ^{+}\pi ^{-},$ but other decay modes were also investigate$.$
The decay of a diquark-antidiquark bound state into a pair of mesons
can occur through the exchange of a quark and an antiquark belonging
respectively to the diquark and the antidiquark. There are indeed
three different flavor configurations.
Thus we need to introduce three amplitudes. Two of them account for $%
X\rightarrow D^{0}\bar{D}^{0\ast }$:

Third one the exchange of a light quark and a heavy quark accounts
for$,$
the charmonium channels i-e,%
\begin{equation}
\text{\textbf{A}}\mathbf{([}cq][\bar{c}\bar{q}]\rightarrow \lbrack q\bar{q}%
][c\bar{c}]\mathbf{)}\equiv \text{A}  \label{2.45}
\end{equation}%
The only available ones are $J/\psi \rightarrow 2\pi $ and $J/\psi
\rightarrow 3\pi $, dominated by $\rho ^{0}$ and $\omega $
respectively and were confirmed experimentaly.

\begin{figure}[here]
\centering
    \includegraphics[scale=.5]{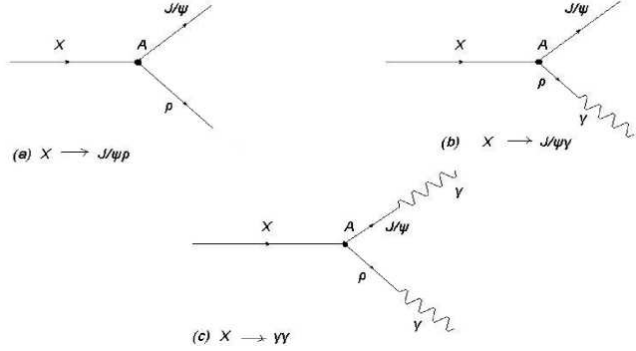}
    \caption{Radiative and Hadronic
decay of the $X(3872)$ described with the same contact vertex A. The
radiative decay proceeds through the hadronic transition.}
\end{figure}

\subsection{Hadronic Decays}

The decay rate for $X\rightarrow J/\psi +f,$ as shown in Figure
3-2(a) can be written as

\begin{equation}
\frac{d\Gamma (X_{[c,l]}\rightarrow \psi +f)}{ds}=\frac{2x_{l,V}\
|A|^{2}B_{(V\rightarrow f)}}{8\pi M_{X_{[c,l]}}^{2}}.\left[ \frac{%
M_{V}\Gamma _{V}}{\pi }\frac{p(s)}{(s-M_{V}^{2})^{2}+(M_{V}\Gamma _{V})^{2}}%
\right]  \label{2.46}
\end{equation}%
with
\[
f=\pi ^{+}\pi ^{-}(\pi ^{+}\pi ^{-}\pi ^{0})\text{ \ for }V=\rho
(\omega )
\]%
$p$ the decay momentum:%
\begin{eqnarray}
p(s) &=&\frac{\sqrt{\lambda (M_{X_{[c,l]}},\ M_{\psi },\ M_{V})}}{%
2M_{X_{[c,l]}}};  \label{2.47} \\
\lambda &=&(M_{X_{[c,l]}})^{4}+(M_{\psi
})^{4}+(M_{V})^{4}-2(M_{X_{[c,l]}}M_{\psi
})^{2}-2(M_{X_{[c,l]}}M_{V})^{2}
\nonumber \\
&&-2(M_{\psi }M_{V})^{2}  \label{2.48}
\end{eqnarray}%
where the coefficient $x_{l,V}\ $is:%
\[
x_{l,V}\ =\frac{(\cos \theta \ \pm \sin \theta )^{2}\;}{2}
\]%
and $A$ is taken to be $2.6$ $GeV\ $\cite{20}. Similarly we can
derive all
above equations for higher mass state $X_{[c,h]}.$ Numerical integration of%
\[
\langle p\rangle _{\rho }=\left( \frac{M_{\rho }\Gamma _{\rho }}{\pi }%
\right) \int_{(2m_{\pi })^{2}}^{\infty }ds\ \frac{p(s)}{(s-M_{\rho
}^{2})^{2}+(M_{\rho }\Gamma _{\rho })^{2}}
\]

\[
\langle p\rangle _{\omega }=\left( \frac{M_{\rho }\Gamma _{\rho }}{\pi }%
\right) \int_{(3m_{\pi })^{2}}^{\infty }ds\ \frac{p(s)}{(s-M_{\omega
}^{2})^{2}+(M_{\omega }\Gamma _{\omega })^{2}}
\]%
gives

\[
\langle p\rangle _{\rho }=126\ MeV,\ \ \ \langle p\rangle _{\omega
}=22\ MeV
\]

\[
\Gamma (X_{[c,l]}\rightarrow \psi +\pi ^{+}\pi
^{-})=\frac{2x_{l,\rho }\ |A|^{2}}{8\pi M_{X_{[c,l]}}^{2}}\langle
p\rangle _{\rho }
\]

\[
\Gamma (X_{[c,l]}\rightarrow \psi +\pi ^{+}\pi ^{-}\pi ^{0})=\frac{%
2x_{l,\omega }\ |A|^{2}}{8\pi M_{X_{[c,l]}}^{2}}\langle p\rangle
_{\omega }
\]

\begin{eqnarray}
\Gamma (X_{[c,l]} &\rightarrow &\psi +\pi ^{+}\pi ^{-})=2x_{l,\rho
}\cdot
2.3\ MeV=3.78\text{ }MeV  \label{2.49} \\
\Gamma (X_{[c,l]} &\rightarrow &\psi +\pi ^{+}\pi ^{-}\pi
^{0})=2x_{l,\omega }\cdot 0.4MeV=0.66\text{ }MeV  \label{2.50}
\end{eqnarray}%
Using these values of decay rates, we can get the information about
the mixing angle $\theta $.

\begin{eqnarray}
\left( \frac{\Gamma (3\pi )}{\Gamma (2\pi )}\right) _{X_{[c,l]}} &=&\frac{%
(\cos \theta \ +\sin \theta )^{2}}{(\cos \theta \ -\sin \theta
)^{2}}\cdot \frac{\langle p\rangle _{\omega }}{\langle p\rangle
_{\rho }}=0.802
\label{2.51} \\
\left( \frac{\Gamma (3\pi )}{\Gamma (2\pi )}\right) _{X_{[c,h]}} &=&\frac{%
(\cos \theta \ -\sin \theta )^{2}}{(\cos \theta \ +\sin \theta
)^{2}}\cdot \frac{\langle p\rangle _{\omega }}{\langle p\rangle
_{\rho }}=0.802 \label{2.52}
\end{eqnarray}%
where from the Belle experiment we have

\[
\left( \frac{\Gamma (3\pi )}{\Gamma (2\pi )}\right) _{Belle}=0.8\pm
0.3_{stat}\pm 0.1_{syst}
\]%
Putting every thing together we have $\theta \ =\pm 20^{0}$ , for
$X_{[c,l]}$ and $X_{[c,h]}$ respectively. For the charged state
$X^{\pm },$ that decay
via $\rho $-exchange only: we have%
\[
\Gamma (X^{\pm }\rightarrow J/\psi \pi ^{\pm }\pi ^{0})=2|A^{2}|\frac{%
\langle p\rangle _{\rho }}{8\pi M_{X}^{2}}=4.6\text{ MeV}
\]

\subsection{Radiative Decays}

The amplitude for the radiative decay $X$ $\rightarrow $ $J/\psi
\gamma $ proceeds through the annihilation of a pair of light quarks
into a photon, the hadronic part of the amplitude is the same as in
the decay $X\rightarrow $ $J/\psi \pi ^{+}\pi ^{-}$. The radiative
decay proceeds through the hadronic transition $X\rightarrow J/\psi
\rho .$ Exploiting the Vector Meson Dominance (VMD), which describe
interactions between photons and hadronic matter \cite{22}, one can
write transition matrix from the Fig. 3-2(b) as:

\begin{equation}
\langle J/\psi \gamma |X\rangle =\langle \gamma |\rho \rangle \frac{1}{%
m_{\rho }^{2}}\langle J/\psi \rho |X\rangle =\frac{f_{\rho
}}{m_{\rho }^{2}}A \label{2.53}
\end{equation}%
Thus the partial decay width is:%
\begin{equation}
\Gamma (X\rightarrow J/\psi \gamma )=2|A^{2}|\left( \frac{f_{\rho
}}{m_{\rho }^{2}}\right) ^{2}\frac{1}{8\pi
M_{X}^{2}}\frac{\sqrt{\lambda (M_{X},\ M_{\psi },\ 0)}}{2M_{X}}
\label{2.54}
\end{equation}%
Using $f_{\rho }=0.152$ GeV$^{2}$ \cite{21} we get,%
\[
\frac{\Gamma (X\rightarrow J/\psi \gamma )}{\Gamma (X\rightarrow
J/\psi \pi ^{+}\pi ^{-})}\sim 0.44
\]%
which is in agreement with experimental value \cite{31}.%
\[
\frac{\Gamma (X\rightarrow J/\psi \gamma )}{\Gamma (X\rightarrow
J/\psi \pi ^{+}\pi ^{-})}<0.40
\]%
Similarly it is easy to calculate other radiative decays. The
experimental
values for these radiative decays\ are \cite{31}:%
\begin{eqnarray*}
\frac{\Gamma (X\rightarrow \chi _{c1}\gamma )}{\Gamma (X\rightarrow
J/\psi
\pi ^{+}\pi ^{-})} &\sim &0.89 \\
\frac{\Gamma (X\rightarrow \chi _{c2}\gamma )}{\Gamma (X\rightarrow
J/\psi \pi ^{+}\pi ^{-})} &\sim &1.1
\end{eqnarray*}%
\ We exploit the result obtained for the width of $X$ $\rightarrow $
$J/\psi
\gamma $\ to give an estimate of the decay width into $J/\psi \gamma \gamma $%
. We compute the transition matrix element in terms of A using the
coupling
of the $J/\psi $ to the $\gamma $ for the Figure 3-2(c)$.$%
\begin{equation}
\langle \gamma \gamma |X\rangle =\langle \gamma |J/\psi \rangle \frac{1}{%
m_{J/\psi }^{2}}\langle J/\psi \gamma |X\rangle =\frac{f_{J/\psi }}{%
m_{J/\psi }^{2}}\langle J/\psi \gamma |X\rangle =\frac{f_{J/\psi }}{%
m_{J/\psi }^{2}}\frac{f_{\rho }}{m_{\rho }^{2}}A  \label{2.55}
\end{equation}%
Thus the partial decay width is:%
\begin{equation}
\Gamma (X\rightarrow \gamma \gamma )=\frac{4}{3}|A^{2}|\left( \frac{%
f_{J/\psi }}{m_{J/\psi }^{2}}\right) ^{2}\left( \frac{f_{\rho
}}{m_{\rho
}^{2}}\right) ^{2}\frac{1}{8\pi M_{X}^{2}}\frac{\sqrt{\lambda (M_{X},\ 0,\ 0)%
}}{2M_{X}}  \label{2.56}
\end{equation}%
Using $f_{J/\psi }=1.254$ GeV$^{2}$ \cite{21} we get,%
\[
\frac{\Gamma (X\rightarrow \gamma \gamma )}{\Gamma (X\rightarrow
J/\psi \pi ^{+}\pi ^{-})}\sim 3\times 10^{-5}
\]%
that is greater than the upper limit provided in experimental data \cite{54}.%
\[
\frac{\Gamma (X\rightarrow \gamma \gamma )}{\Gamma (X\rightarrow
J/\psi \pi ^{+}\pi ^{-})}<1.5\times 10^{-5}
\]%
The inconsistency of the theoretical prediction with respect to data
is not dramatic if we take into account the very strong assumptions
made to derive the Eq.(\ref{2.56}).

\chapter{GROWING EVIDENCE OF TETRAQUARKS: THE $Y_{b}(10890)$}

\section{Introduction}

Using the diquark-antidiquark model, in the previous chapter we
learnt to
derive the spectrum for lowest lying $[cq][\bar{c}\bar{q}]$ states with $q=u$%
, $d,$ and discuss the decays of $X(3872)$. This model can be easily
applied to lowest lying $[bq][\bar{b}\bar{q}]$ states with $q=u$,
$d$, $s$, and $c.$
In this chapter we will\ discuss the spectrum of the higher mass $[bq][\bar{b%
}\bar{q}]$ tetraquark states with $q=u$, $d.$ Particularly we are
interested in the decays of higher tetraquark particles with $q=u$,
$d$. First,\ there is evidence for $s\bar{s}$ bound state,
$Y_{s}(2175)$ having the quantum numbers $J^{PC}=1^{--}$, first
observed by BaBar in the initial state
radiation $(ISR)$ process $e^{+}e^{-}\rightarrow \gamma _{\mathrm{ISR}%
}\;f_{0}(980),$ $\phi (1020)$, where $f_{0}(980)$ is the $0^{++}$
scalar
state~\cite{32}. This was later confirmed by BES~\cite{33} and Belle~\cite%
{34}. In December $2007$, the Belle collaboration working at the KEKB $%
e^{+}e^{-}$ collider in Tsukuba, Japan, reported the first
observation of the processes $e^{+}e^{-}\rightarrow
Y_{[bq]}\rightarrow \Upsilon (1S,2S)\;\pi ^{+}\pi ^{-}$ near the
peak of the $\Upsilon (5S)$ resonance at the center-of-mass energy
of about $10.87$ $GeV$ \cite{41}. Belle measurements near the
$\Upsilon (5S)$, however, did not fall in line with theoretical
expectations \cite{41}. Their data were enigmatic in the partial
decay widths for $\Upsilon (5S)$ $\rightarrow \Upsilon (1S)$ $\pi
^{+}\pi ^{-}$\ and $\Upsilon (2S)$ $\pi ^{+}\pi ^{-}$\ which were
typically three orders of magnitude larger than anticipated in QCD
\cite{42}. Production and decays of the $\Upsilon (nS)$ states ($n$
being the principal quantum number) are popular theoretical
laboratories to test QCD. In particular, the final states $\Upsilon
(1S,2S)\;\pi ^{+}\pi ^{-}$ arising from the production and decays of
the lower bottomonia states, such as $\Upsilon (4S)$ $\rightarrow $
$\Upsilon (1S)$ $\pi ^{+}\pi ^{-}$, have been studied in a number of
experiments over the last thirty years and are theoretically
well-understood in QCD \cite{42}. In addition, the dipion invariant
mass distributions in these events were distinctly different from
theoretical expectations as well as from the corresponding
measurements at the $\Upsilon (4S)$, undertaken previously by Belle.

In the conventional Quarkonium theory, there is no place for such a
nearby additional $b\bar{b}$ resonance having the quantum numbers of
$\Upsilon (5S). $ An important issue is whether the puzzling events
seen by Belle stem from the decays of the $\Upsilon (5S)$, or from
another particle $Y_{b}$ having a mass close enough to the mass of
the $\Upsilon (5S)$. We will see that the interpretation of the
Belle data is that the anomalous $\Upsilon (1S,2S)\;\pi ^{+}\pi
^{-}$ events are not due to the production and decays of the
$\Upsilon (5S)$, but rather from the production of a completely
different hadron species, tetraquark hadrons with the quark structure $%
Y_{[bq]}=[bq][\bar{b}\bar{q}]$ states with $q=u$, $d$ and their
subsequent decays. A. Ali et. al. call this state a "Brand New Form
of Matter". Identifying the $J^{PC}=1^{--}$ state $Y_{[bq]}(10900)$
seen in the energy scan of the $e^{+}e^{-}\rightarrow b\bar{b}$
cross section by BaBar~\cite{30} with the state $Y_{[bq]}(10890)$
seen by Belle~\cite{41}.

Clearly, two aspects of the Belle data had to be explained:

(a) the anomalously large partial decay rates and

(b) the invariant mass distributions of the dipions.

A dynamical model based on the tetraquark interpretation of
$Y_{[bq]}(10890)$ was presented in \cite{49} where it was pointed
that it is in agreement with the measured distributions in the
decays $Y_{[bq]}\rightarrow \Upsilon (1S)\;\pi ^{+}\pi ^{-},\Upsilon
(2S)\;\pi ^{+}\pi ^{-}$. They have argued that the decays
$Y_{[bq]}\rightarrow \Upsilon (1S,2S)\pi ^{+}\pi ^{-}$ are
radically different than the similar dipion transitions measured in the $%
\Upsilon (4S)$ and lower mass Quarkonia. Most resently the decays $%
Y_{[bq]}\rightarrow \Upsilon (1S)K^{+}K^{-}$ was investigated by
A.~Ali \textit{et al. }\cite{50} further supporting the\ Belle
data$.$ We will discuss in detail the anomalously large partial
decay rates and not the invariant mass distributions of the dipions.

\section{The Fifth Quark Flavor: Bottom Mesons}

Fifth quark was discovered, when in $1977$ the upsilon meson \thinspace $%
\Upsilon (J^{PC}=1^{--})$ was found experimentally as a narrow
resonance at Fermi Lab. with mass $\sim 9.5$ GeV. This was later
confirmed in $e^{+}e^{-}$ experiments at DESY and CESR which
determined its mass to be $9460\pm 10$
MeV and also its width. The updated parameters of this resonance are mass $%
9460.37\pm 0.21$ MeV and width $52.5\pm 1.8$ keV. Again the narrow
width in spite of large phase space available suggests the existence
of a fifth quark flavor called beauty, with a new quantum number
$B=-1$ for the bottom $(b)$ quark. With this assignment the formula
$Q=I_{3}+1/2(Y+B+C)$ would give the charge of $b$ quark the value
$-1/3(I_{3}=0)$. The mass of $b$ quark is expected to be around
$4.9$ GeV as suggested by the $\Upsilon $ mass which is regarded as
a $^{3}S_{1}$ bound state of $b\bar{b}$.

One would also expect the particles with $B=\pm 1$, such as $b\bar{q}$ or $q%
\bar{b}$. The lowest lying bound states $b\bar{q}$ and $q\bar{b}$
have been
found experimentally. The $B=-1$ states $(\bar{B}^{0},\,B^{-})\bar{B}%
_{s}^{0} $ form an SU(3) triplet $(\bar{3})$ and $B=+1$ states $({B}%
^{+},\,B^{0})B_{s}^{0}$ form another triplet $(3)$. For p-wave
multiplets

\begin{tabular}{lll}
$(q\bar{b})_{L=1}$ & $J^{P}=2^{+},1^{+}$ & $\left.
\begin{tabular}{l}
$({B}_{2}^{\ast +,0},\,B_{1}^{+,0})$ \\
$({B}_{s_{2}}^{\ast 0},{B}_{s_{1}}^{\ast 0}\,)$%
\end{tabular}%
\right] _{j=3/2}$ \\
& $J^{P}=1^{+},0^{+}$ & $\left.
\begin{tabular}{l}
$({B}_{1}^{\ast +,0},\,B_{0}^{\ast +,0})$ \\
$({B}_{s_{1}}^{\ast 0},{B}_{s_{0}}^{\ast 0})$%
\end{tabular}%
\right] _{j=1/2}$%
\end{tabular}

The masses and decay time of B-mesons are given below%
\begin{eqnarray*}
{B}^{\pm } &=&5279.16\pm 0.31\text{ }MeV,\text{ \ }\tau =(1.638\pm
0.11)\times 10^{-12}\sec \\
{B}^{0} &=&5279.53\pm 0.33\text{ }MeV,\text{ \ }\tau =(1.530\pm
0.069)\times 10^{-12}\sec
\end{eqnarray*}

\section{Spectrum Of Higher Mass $[bq][\bar{b}\bar{q}]$ States}

For the orbital excitation $L_{Q\bar{Q}}=1$ we consider both good
and bad diquarks. The orbital excitation $L_{Q\bar{Q}}=1$ leads to
negative parity states,where $1^{--}$ multiplet is of main interest
in this chapter. To
estimate the masses, we repeat the diagonalization with the basis:%
\begin{eqnarray}
\left\vert 1\right\rangle &=&\left\vert 0_{\mathcal{Q}},0_{\bar{\mathcal{Q}}%
};~1_{J}\right\rangle  \nonumber \\
\left\vert 2\right\rangle &=&\frac{1}{\sqrt{2}}\left( \left\vert 0_{\mathcal{%
Q}},1_{\bar{\mathcal{Q}}};~1_{J}\right\rangle +\left\vert 1_{\mathcal{Q}},0_{%
\bar{\mathcal{Q}}};~1_{J}\right\rangle \right)  \nonumber \\
\left\vert 3\right\rangle &=&\left\vert 1_{\mathcal{Q}},1_{\bar{\mathcal{Q}}%
};~1_{J}\right\rangle  \label{3.1}
\end{eqnarray}%
Since for the both good and bad diquarks parity is positive as from Eq.(\ref%
{1.8}) and Eq.(\ref{1.9}) ($0^{+}$ and $1^{+}$ respectively), the state $%
\left\vert 2\right\rangle $ has $P=C=-1$, provided that
$L_{Q\bar{Q}}=1.$ Since
\[
C_{Q\bar{Q}}(-1)^{L_{Q\bar{Q}}}(-1)^{S_{Q\bar{Q}}}=1
\]%
therefore for the states $\left\vert 1\right\rangle $\ and
$\left\vert
3\right\rangle ,$ $C_{Q\bar{Q}}=-1$ provided that $S_{Q\bar{Q}}=0,$ $2$ and $%
L_{Q\bar{Q}}=1.$ First of all we have to change the Hamiltonian given in Eq.(%
\ref{2.16}) by replacing the charm quark by bottom quark.%
\begin{equation}
H=2m_{\mathcal{Q}}+H_{SS}^{(\mathcal{QQ)}}+H_{SS}^{(\mathcal{Q\bar{Q}%
\mathcal{)}}}+H_{SL}+H_{LL}  \label{3.2}
\end{equation}%
where:%
\begin{eqnarray}
H_{SS}^{(\mathcal{QQ)}} &=&2(\mathcal{K}_{bq})_{\bar{3}}[(\mathbf{S}%
_{b}\cdot \mathbf{S}_{q})+(\mathbf{S}_{\bar{b}}\cdot
\mathbf{S}_{\bar{q}})],
\label{3.2(a)} \\
H_{SS}^{(\mathcal{Q\bar{Q}\mathcal{)}}} &=&2(\mathcal{K}_{b\bar{q}})(\mathbf{%
S}_{b}\cdot \mathbf{S}_{\bar{q}}+\mathbf{S}_{\bar{b}}\cdot \mathbf{S}_{q})+2%
\mathcal{K}_{b\bar{b}}(\mathbf{S}_{b}\cdot \mathbf{S}_{\bar{b}})+2\mathcal{K}%
_{q\bar{q}}(\mathbf{S}_{q}\cdot \mathbf{S}_{\bar{q}}),  \label{3.2(b)} \\
H_{SL} &=&2A_{\mathcal{Q}}(\mathbf{S}_{\mathcal{Q}}\cdot \mathbf{L}+\mathbf{S%
}_{\mathcal{\bar{Q}}}\cdot \mathbf{L}),  \label{3.2(c)} \\
H_{LL} &=&B_{\mathcal{Q}}\frac{L(L+1)}{2}.  \label{3.2(d)}
\end{eqnarray}%
To perform the digonalization we use the same shorthand notation as
described in the previous chapter (c.f. Eq.(\ref{2.17}) to
Eq.(\ref{2.26})) for the basis vectors defined in Eq.(\ref{3.1}).\
We derive the mass term shift $\Delta m_{SS}$\ for higher mass
$[bq][\bar{b}\bar{q}]$ states, due to
the part of the Hamiltonian containing only spin-spin interaction terms, $%
H_{SS}.$ Let us first consider
\[
(H_{SS})_{33}=\langle 1_{\mathcal{Q}},1_{\bar{\mathcal{Q}}};~1_{J}|(H_{SS}^{(%
\mathcal{QQ)}}+H_{SS}^{(\mathcal{Q\bar{Q}\mathcal{)}}})|1_{\mathcal{Q}},1_{%
\bar{\mathcal{Q}}};~1_{J}\rangle
\]%
( Here we use $\frac{1}{2}[s_{\mathcal{Q}}(s_{\mathcal{Q}}+1)-\frac{3}{2}]=%
\vec{S}_{\mathcal{Q}}.\vec{S}_{\mathcal{Q}}$)%
\begin{eqnarray}
(H_{SS})_{33} &=&\langle 1_{\mathcal{Q}},1_{\bar{\mathcal{Q}}%
};~1_{J}|H_{SS}^{(\mathcal{QQ)}}|1_{\mathcal{Q}},1_{\bar{\mathcal{Q}}%
};~1_{J}\rangle +\langle 1_{\mathcal{Q}},1_{\bar{\mathcal{Q}}%
};~1_{J}|H_{SS}^{(\mathcal{Q\bar{Q}\mathcal{)}}}|1_{\mathcal{Q}},1_{\bar{%
\mathcal{Q}}};~1_{J}\rangle  \label{3.3} \\
&=&2\left( \mathcal{K}_{bq}\right) _{\bar{3}}(2-3/2)+2\mathcal{K}_{b\bar{q}%
}[(2-3/2)\frac{1}{2}+(2-3/2)\frac{1}{2}]  \nonumber \\
&&+2\mathcal{K}_{q\bar{q}}(2-3/2)\frac{1}{2}+2\mathcal{K}_{b\bar{b}}(2-3/2)%
\frac{1}{2}  \nonumber \\
&=&\left( \mathcal{K}_{bq}\right) _{\bar{3}}-\mathcal{K}_{b\bar{q}}-\frac{1}{%
2}\mathcal{K}_{q\bar{q}}-\frac{1}{2}\mathcal{K}_{b\bar{b}}
\label{3.4}
\end{eqnarray}%
Similarly we can easily calculate $(H_{SS})_{11}$ and
$(H_{SS})_{22}.$ All off diagonal elements are zero. Thus finally we
have,

\begin{equation}
\Delta m_{SS}=\left(
\begin{array}{ccc}
-3\left( \mathcal{K}_{bq}\right) _{\bar{3}} & 0 & 0 \\
0 &
\begin{array}{c}
-\left( \mathcal{K}_{bq}\right) _{\bar{3}}-\mathcal{K}_{b\bar{q}} \\
+\left( \mathcal{K}_{q\bar{q}}+\mathcal{K}_{b\bar{b}}\right) /2%
\end{array}
& 0 \\
0 & 0 &
\begin{array}{c}
\left( \mathcal{K}_{bq}\right) _{\bar{3}}-\mathcal{K}_{b\bar{q}} \\
-\left( \mathcal{K}_{q\bar{q}}+\mathcal{K}_{b\bar{b}}\right) /2%
\end{array}%
\end{array}%
\right)  \label{3.5}
\end{equation}%
The eigenvalues of the spin-orbit and angular momentum operators
given in
Eq.(\ref{3.2(c)}) and Eq.(\ref{3.2(d)}), were calculated by Polosa et al.~%
\cite{17}, we have

\begin{eqnarray*}
A_{\mathcal{Q}} &=&5\text{ MeV, for }q=u\text{, }d, \\
B_{\mathcal{Q}} &=&408\text{ MeV, for }q=u\text{, }d,
\end{eqnarray*}%
We use these values in order to calculate the numerical values of
these
states. Hence the eight tetraquark states $[bq][\bar{b}\bar{q}]$ ($q=u,$ $d$%
) having the quantum numbers $1^{--}$ are:%
\begin{eqnarray}
M_{Y_{[bq]}}^{(1)}\left( S_{\mathcal{Q}}=0,~S_{\bar{\mathcal{Q}}}=0,~S_{%
\mathcal{Q\bar{Q}}}=0,~L_{\mathcal{Q\bar{Q}}}=1\right) &=&2m_{\left[ bq%
\right] }+\lambda _{1}+B_{\mathcal{Q}},  \label{3.6} \\
M_{Y_{[bq]}}^{(2)}\left( S_{\mathcal{Q}}=1,~S_{\bar{\mathcal{Q}}}=0,~S_{%
\mathcal{Q\bar{Q}}}=1,~L_{\mathcal{Q\bar{Q}}}=1\right) &=&2m_{\left[ bq%
\right] }+\Delta +\lambda _{2}-2A_{Q}+B_{\mathcal{Q}},  \label{3.7} \\
M_{Y_{[bq]}}^{(3)}\left( S_{Q}=1,~S_{\bar{Q}}=1,~S_{Q\bar{Q}}=0,~L_{Q\bar{Q}%
}=1\right) &=&2m_{\left[ bq\right] }+2\Delta +\lambda
_{3}+B_{\mathcal{Q}},
\label{3.8} \\
M_{Y_{[bq]}}^{(4)}\left( S_{Q}=1,~S_{\bar{Q}}=1,~S_{Q\bar{Q}}=2,~L_{Q\bar{Q}%
}=1\right) &=&2m_{\left[ bq\right] }+2\Delta +\lambda _{3}-6A_{Q}+B_{%
\mathcal{Q}},  \label{3.9}
\end{eqnarray}%
where $\lambda $'s are the diagonal elements of the matrix $\Delta
M_{SS}$ given in Eq.(\ref{3.5}). The quantity $\Delta ,$ is the mass
difference of
the \textit{good} and the \textit{bad} diquarks i.e.%
\[
\Delta =m_{\mathcal{Q}}\left( S_{\mathcal{Q}}=1\right) -m_{\mathcal{Q}%
}\left( S_{\mathcal{Q}}=0\right) .
\]%
and from Eq.(\ref{2.38})%
\[
m_{[bq]}=m_{[cq]}-m_{c}+m_{b}
\]%
Now one of the remaining unknowns in this calculation is the quantity $%
\Delta $, the mass difference of the \textit{good} and the
\textit{bad} diquarks. Following Jaffe and Wilczek \cite{7}, the
value of $\Delta $ for diquark $[bq]$ is $202$ MeV.

We recall previous chapter where we have used the known mesons and
baryons to calculate the couplings of the spin-spin interaction. We
can extend the
same procedure to the $S=1$, $L=(0$, $1$) meson states $B^{\ast }$, $%
B_{1}\left( 5721\right) $, $B_{2}\left( 5747\right) $ to calculate
the values:

\begin{table}[tb]
\caption{Spin-Spin couplings for quark-antiquark pairs in in the
color singlet state from the known bottom mesons.}
\begin{center}
\begin{tabular}{|l|l|l|l|}
\hline Spin-spin couplings & $q\bar{q}$ & $b\bar{q}$ & $b\bar{b}$ \\
\hline
$\left( \mathcal{K}_{ij}\right) _{0}$(MeV) & $318$ & $23$ & $36$ \\
\hline\hline
\end{tabular}%
\end{center}
\end{table}
\begin{table}[tb]
\caption{Spin-Spin couplings for quark-quark in color $\bar{3}$
state from the known bottom baryons.}
\begin{center}
\begin{tabular}{|l|l|l|}
\hline Spin-Spin couplings & $qq$ & $bq$ \\ \hline $\left(
\mathcal{K}_{ij}\right) _{\bar{3}}$(MeV) & $98$ & $6$ \\
\hline\hline
\end{tabular}%
\end{center}
\end{table}

Ultimately by putting things together, the masses for the states
given in Eqs.(\ref{3.6}-\ref{3.9}) are given in Table and can be
compared with the ones estimated in refs. \cite{36} using the QCD
sum rules.

\begin{table}[tb]
\caption{Masses of the the $1^{--}$ neutral tetraquark states $M_{{Y}%
_{[bq]}}^{(n)}$ in GeV. The value $M_{{Y}_{[bq]}^{(1)}}$ (for
$q=u,d$) is fixed to be 10.890 GeV, identifying this with the mass
of the $Y_{b}$ from BELLE.~} .
\par
\begin{center}
$%
\begin{tabular}{||l||l||l||l||}
\hline\hline
$M_{Y_{[bq]}}^{(i)}$ & $M_{Y_{[bq]}}^{(1)}$ & $M_{Y_{[bq]}}^{(2)}$ & $%
M_{Y_{[bq]}}^{(3)}$ \\ \hline\hline $q=u,d$ & $10890$ & $11130$ &
$11257$ \\ \hline\hline
\end{tabular}%
\begin{tabular}{|l||}
\hline\hline $M_{Y_{[bq]}}^{(4)}$ \\ \hline\hline $11227$ \\
\hline\hline
\end{tabular}%
$%
\end{center}
\end{table}

Note that there are 8 electrically neutral self-conjugate $1^{--}$
tetraquark states $Y_{[bq]}^{(n)}$ with the quark contents $[bq][\bar{b}\bar{%
q}]$, with $q=u$, $d$ of which the two corresponding to $[bu][\bar{b}\bar{u}%
] $ and $[bd][\bar{b}\bar{d}]$, i.e., $Y_{[bu]}^{(n)}$ and
$Y_{[bd]}^{(n)}$ are degenerate in mass due to the isospin symmetry.
Their mass difference is induced by isospin splitting $m_{d}-m_{u}$,
mixing angle $\theta $ and is estimated as $\Delta M(Y_{b})=(5.6\pm
2.8)$ MeV. Due to this small differnce in the following we will not
distinguish between the lighter and the heavier of these states and
denote them by the common symbol $Y_{b}.$ There are yet more
electrically neutral $J^{PC}=1^{--}$ states with the mixed light
quark
content $[bd][\bar{b}\bar{s}]$ and their charge conjugates $[bs][\bar{b}\bar{%
d}]$. However, these mixed states don't couple directly to the photons, $%
Z^{0}$ or the gluon, and are not of our main interest.

\section{Decay Widths Of The $Y_{b}(10890)$}

\subsection{Leptonic Decay Widths}

For bottomonium systems, the corresponding decay widths are
determined by
the wave functions at the origin for the $\Upsilon (nS)$, $\Psi _{b\bar{b}%
}(0)$, and by the derivative of these functions at the origin, $\Psi _{b\bar{%
b}}^{\prime }(0)$, for the P-waves. To take into account the
possibly larger hadronic size of the tetraquarks compared to that of
the $b\bar{b}$ mesons, we modify the Quarkonia potential, usually
taken as a sum of linear
(confining) and Coulombic (short-distance) parts. For example, the Buchm\"{u}%
ller-Tye $Q\bar{Q}$ potential~\cite{38} has the asymptotic forms

\begin{eqnarray}
V(r) &\sim &k_{Q\bar{Q}}\;r\text{ },\text{\ \ \ \ \ \ \ \ \ \ \ \ \
\ \ \ \
\ \ \ \ \ }(\text{for }r\rightarrow \infty )  \label{3.10} \\
V(r) &\sim &1/r\ln (1/\Lambda _{\mathrm{QCD}}^{2}\;r^{2})\text{
},\text{\ \ \ \ \ \ }(\text{for }r\rightarrow 0)  \label{3.11}
\end{eqnarray}%
where $k_{Q\bar{Q}}$ is the string tension and $\Lambda
_{\mathrm{QCD}}$ is
the QCD scale parameter. The bound state tetraquark potential $V_{\mathcal{Q%
\bar{Q}}}(r)$ will differ from the Quarkonia potential
$V_{Q\bar{Q}}(r)$ in the linear part, as the string tension in a
diquark $k_{\mathcal{QQ}}$ is expected to be different than the
corresponding string tension $k_{Q\bar{Q}}$ in the $Q\bar{Q}$
mesons. The diquarks-antidiquarks in the tetraquarks and the
quarks-antiquarks in the mesons are in the same $\bar{3}_{c}3_{c}$
color configuration, the Coulomb (short-distance) parts of the
potentials will be similar. Defining,

\begin{equation}
\kappa =k_{\mathcal{Q\bar{Q}}}/k_{Q\bar{Q}}  \label{3.12}
\end{equation}%
we expect $\kappa $ to have a value in the range $\kappa \in \lbrack \frac{1%
}{2},\frac{\sqrt{3}}{2}]$ $\cite{39}$. This will modify the
tetraquark wave functions $\Psi _{\mathcal{Q\bar{Q}}}(0)$ from the
corresponding wave functions of the bound $b\bar{b}$ systems,
effecting the decay amplitudes and hence all the decay widths of the
tetraquarks. The corresponding value for the tetraquark states
$[bq][\bar{b}\bar{q}]$ is then calculated taking into account the
ratio of the string tensions $\kappa $. As the linear part of the
confining potential determines essentially the heavy Quarkonia wave
functions, we find that to a good approximation:
\begin{equation}
\Psi _{\mathcal{Q\bar{Q}}}(0)\simeq \kappa \Psi _{b\bar{b}}(0)
\label{3.13}
\end{equation}%
which is what we can use in our derivations of the decay widths.

The partial electronic widths $\Gamma _{ee}(Y_{[b,l]})$ and $\Gamma
_{ee}(Y_{[b,h]})$ are given by the well known Van Royen-Weisskopf
formula for the P-states, which we write as:
\begin{equation}
\Gamma _{ee}=\frac{16\pi Q^{2}\alpha ^{2}|\Psi
_{\mathcal{Q\bar{Q}}}^{\prime }|^{2}}{M^{2}\omega ^{2}},
\label{3.14}
\end{equation}%
where $Q=-2/3$ is the diquark charge in $Y_{bd}=[bd][\bar{b}\bar{d}]$ and $%
Q=+1/3$ is the charge of the diquarks in $Y_{bu}=[bu][\bar{b}\bar{u}]$, $%
\alpha =1/137$ is the electromagnetic coupling constant to lowest order and $%
\Psi _{\mathcal{Q\bar{Q}}}^{\prime }(\vec{r})=\psi (\phi ,\theta
)R^{\prime }(r)$ is the first derivative in $r$ of the wave function
of the tetraquark, which needs to be taken at the origin i.e.,

\begin{equation}
\Psi _{\mathcal{Q\bar{Q}}}^{\prime }(\vec{r})=\sqrt{\frac{3}{4\pi }}%
R^{\prime }(0)  \label{3.15}
\end{equation}%
The wave function $\Psi _{b\bar{b}}^{\prime }(0)$ depends on the
underlying potential model. Tetraquarks are bound state of
diquarks-antidiquarks with
the same internal color structure as the Quarkonia, i.e. both are $(\mathbf{%
3}$, $\mathbf{\bar{3}})$ bound states. Hence, they also depend, in
principle, on the potential models. We will not distinguish between
the lighter and the heavier of these states because\ their mass
difference is induced by isospin splitting $m_{d}-m_{u}$ and a
mixing angle and is estimated as $\Delta M(Y_{b})=(5.6\pm 2.8)$ MeV.
So we denote them by the common symbol $Y_{b}.$

To calculate the radial wave function, we took the value calculated
by using the QQ-onia package \cite{40} yielding $\left\vert
R^{\prime }(0)\right\vert ^{2}=2.062GeV$. Since each derivative
increases the energy dimension by and thus we need to normalize by
kinetic energy $\omega \approx m_{\mathcal{Q}}$ of the diquark. For
the lowest lying $1^{--}$ state we get,

\begin{equation}
\Gamma _{ee}\approx 0.12KeV  \label{3.16}
\end{equation}%
Since all the $1^{--}$ states are P-waves, the $R^{\prime }(0)$
value will not change because the masses of the diquarks remain the
same. So the value of $\Gamma _{ee}$ only varies with the mass and
therefore does not change
significantly. This value is close to the experimental value given in \cite%
{30}.

\subsection{Hadronic Decays}

We discuss the two-body hadronic decays of the $Y_{b}(10890)$, $%
Y_{b}(q)\rightarrow B_{q}^{(\ast )}(k)\bar{B}_{q}^{(\ast )}(l)$ .
The decays $Y_{b}\rightarrow \Upsilon (1S,2S)\;\pi ^{+}\pi ^{-}$ are
also Zweig allowed. These decays $Y_{[bq]}\rightarrow \Upsilon
(1S,2S)\;\pi ^{+}\pi ^{-} $ are too much phase space suppressed and
require a dynamical model, which we will also discuss later.

The Vertices of the dominant two-body hadronic decays of the
$Y_{b}(10890)$ are \cite{46}:
\begin{equation}
\begin{tabular}{lll}
$Y_{b}\rightarrow B\bar{B}$ & $\widehat{=}$ & $F(k^{\mu }-l^{\mu })$ \\
&  &  \\
$Y_{b}\rightarrow B\bar{B}^{\ast }$ & $\widehat{=}$ &
$\frac{F}{M}\epsilon
^{\mu \nu \rho \sigma }k_{\rho }l_{\sigma }$ \\
&  &  \\
$Y_{b}\rightarrow B^{\ast }\bar{B}^{\ast }$ & $\widehat{=}$ &
$F(g^{\mu \rho
}(q+l)^{\nu }-g^{\mu \nu }(k+q)^{\rho }+g^{\rho \nu }(q+k)^{\mu })$%
\end{tabular}
\label{3.17}
\end{equation}%
The corresponding decay widths are respectively:%
\begin{equation}
\begin{tabular}{lll}
$\;\Longrightarrow \;\;\Gamma \;$ & $=$ & $\frac{F^{2}|\vec{k}|^{3}}{%
2M^{2}\pi }$ \\
$\;\Longrightarrow \;\;\Gamma \;$ & $=$ & $\frac{F^{2}|\vec{k}|^{3}}{%
4M^{2}\pi }$ \\
$\;\Longrightarrow \;\;\Gamma \;$ & $=$ & $\frac{F^{2}|\vec{k}|^{3}(48|\vec{k%
}|^{4}-104M^{2}|\vec{k}|^{2}+27M^{4})}{2\pi (M^{3}-4|\vec{k}|^{2}M)^{2}}$%
\end{tabular}
\label{3.18}
\end{equation}%
The decaying momentum $|\vec{k}|$ is as given in Eq.(\ref{2.47})
\begin{equation}
|\vec{k}|=\frac{\sqrt{M^{2}-(M_{k}+M_{l})^{2}}\sqrt{M^{2}-(M_{k}-M_{l})^{2}}%
}{2M},  \label{3.19}
\end{equation}%
where $M$ is the mass of the decaying particle and $M_{k}$, $M_{l}$
are the masses of the decay products. The matrix elements are
obtained by
multiplying the vertices by the polarization vectors $\varepsilon ^{(i)\mu }$%
. The polarisation vectors $\varepsilon ^{(i)\mu }$ satisfy the
transversality condition with the polarisation sum
\begin{equation}
\sum\limits_{i}^{3}\varepsilon ^{(i)\mu }(p)\varepsilon ^{(i)\nu
}(p)=-g^{\mu \nu }+\frac{p^{\mu }p^{\nu }}{p^{2}}  \label{3.20}
\end{equation}%
Thus, for the decay $Y_{[bq]}\rightarrow B_{q}\bar{B}_{q},$ the
Lorentz-invariant matrix element is:
\begin{equation}
\mathcal{M}=\varepsilon _{\mu }^{[bq]}F(k^{\mu }-l^{\mu })
\label{3.21}
\end{equation}%
The decay constants $F$ are non-perturbative quantities, which are
beyond the scope in our approximation. We estimate them using the
known two-body decays of $\Upsilon (5S)$, which are described by the
same vertices as given Eq.(\ref{3.17}) \cite{3}. We use the decay
widths for the decays $\Upsilon (5S)\rightarrow
B\bar{B},B\bar{B}^{\ast },B^{\ast }\bar{B}^{\ast }$ from the
PDG, yielding the coupling constants, called $F_{\mathrm{PDG}}$ and $|\vec{k}%
|.$

\begin{table}[h]
\caption{2-body decays $\Upsilon (5S)\rightarrow B^{(\ast
)}\bar{B}^{(\ast )} $, which we use as a reference, with the mass
and the decay widths taken from PDG. The extracted values of the
coupling constants $F$ and the centre of mass momentum $|\vec{k}|$
are also shown.}
\begin{center}
\begin{tabular}{||l||l||l||l||}
\hline\hline Process & $\Gamma _{PDG}\left[ MeV\right] $ & $F_{PDG}$
& $|\vec{k}|\left[ GeV\right] $ \\ \hline\hline
$\Upsilon (5S)\rightarrow B\bar{B}$ & $<13.2$ & $<2.15$ & $1.3$ \\
\hline\hline
$\Upsilon (5S)\rightarrow B\bar{B}^{\ast }$ & $15.4_{-6.6}^{+6.6}$ & $%
3.7_{-0.9}^{+0.7}$ & $1.2$ \\ \hline\hline
$\Upsilon (5S)\rightarrow B^{\ast }\bar{B}^{\ast }$ & $48_{-11}^{+11}$ & $%
1_{-0.12}^{+0.13}$ & $1.0$ \\ \hline\hline
\end{tabular}%
\end{center}
\end{table}
The different hadronic sizes of the $b\bar{b}$ Onia states and the
tetraquarks $Y_{[bq]}$ are taken into account by the quantity
$\kappa $, discussed earlier.

\begin{table}[h]
\caption{ Reduced partial decay widths for the tetraquarks
$Y_{[bq]}^{(i)}$, the extracted value of the coupling constant $F$
and the centre of mass momentum $|\vec{k}|$. The errors in the
entries correspond to the errors in the decay widths in Table 4-5.}
\begin{center}
\begin{tabular}{|l|l|l|l|}
\hline
\text{Decay Mode} & $\Gamma /\kappa ^{2}[MeV]$ & $F$ & $|\vec{k}|[GeV]$ \\
\hline $Y_{[bq]}^{(1)}\rightarrow B\bar{B}$ & $<15$ & $2.15$ & $1.3$
\\ \hline
$Y_{[bq]}^{(1)}\rightarrow BB^{\ast }$ & $18_{-8}^{+8}$ & $3.7$ & $1.2$ \\
\hline $Y_{[bq]}^{(1)}\rightarrow B^{\ast }\bar{B}^{\ast }$ &
$56_{-14}^{+14}$ & $1$ & $1.1$ \\ \hline\hline
\end{tabular}%
\\[0pt]
\vspace{1cm}
\end{center}
\end{table}
Thus, the decay widths of $Y_{[bq]}^{(1)}$ are consistent with the
corresponding measurements by Belle. The other three higher $1^{--}$
states have much larger decay widths and will be correspondingly
more difficult to find.

\subsection{Dynamical Model For $Y_{b}\rightarrow \Upsilon (1S,2S)\;\protect%
\pi ^{+}\protect\pi ^{-}$}

\subsubsection{(Introduction and results)}

Explaining the larger decay rates for the transitions
$Y_{b}(q)\rightarrow \Upsilon (p)+\pi ^{+}(k_{1})+\pi ^{-}(k_{2})$
was not so difficult, as the decays of $Y_{b}$ involve a
recombination of the initial four quarks, as
exemplified below by the process $Y_{[bu]}$ $=$ $[bu][\bar{b}\bar{u}%
]\rightarrow (b\bar{b})(u\bar{u}),$ with the subsequent projection $(b\bar{b}%
)\rightarrow \Upsilon (1S)$ and $(u\bar{u})\rightarrow \pi ^{+}\pi ^{-}.$%
Such quark recombination processes do not require the emission and
absorption of gluons, and are appropriately called Zweig-allowed,
after the co-discoverer of the quark-model, George Zweig. The
relevant diagrams for
the decays $Y_{b}(q)\rightarrow \Upsilon (p)+\pi ^{+}(k_{1})+\pi ^{-}(k_{2})$%
.

The decays $Y_{b}\rightarrow \Upsilon (1S)\;\pi ^{+}\pi
^{-},\Upsilon (2S)\;\pi ^{+}\pi ^{-}$ are sub-dominant, but
Zweig-allowed and involve
essentially the quark rearrangements shown below. With the $J^{PC}$ of the $%
Y_{b}$ and $\Upsilon (nS)$ both $1^{--}$, the $\pi ^{+}\pi ^{-}$
states in the decays $Y_{b}\rightarrow \Upsilon (1S)\;\pi ^{+}\pi
^{-},\Upsilon (2S)\;\pi ^{+}\pi ^{-}$ are allowed to have the
$0^{++}$ and $2^{++}$ quantum numbers. There are only three
low-lying states in the Particle Data
Group (PDG)~ which can contribute as intermediate states, namely the two $%
0^{++}$ states, $f_{0}(600)$ and $f_{0}(980)$, which,
following~\cite{44, 48} we take as the lowest tetraquark states, and
the $2^{++}$ $q\bar{q}$-meson state $f_{2}(1270)$, all of which
decay dominantly into $\pi \pi .$ For the decay $Y_{b}\rightarrow
\Upsilon (1S)\;\pi ^{+}\pi ^{-}$, all three states
contribute. However, kinematics allows only the $f_{0}(600)$ in the decay $%
Y_{b}\rightarrow \Upsilon (2S)\;\pi ^{+}\pi ^{-}$. This model
encodes all these features. Finally the fits of the Belle data with
this model, yield the measured partial decay widths $\Gamma
_{\Upsilon (1S)+2\pi }=0.59\pm
0.04\pm 0.09~${MeV }and $\Gamma _{\Upsilon (2S)+2\pi }=0.85\pm 0.07\pm 0.16~$%
{MeV }\cite{49}.

\subsection{Radiative Decays}

The amplitude for the radiative decay $Y_{b}\rightarrow $ $\chi
_{b}\gamma $ will\ proceed through the annihilation of a pair of
light quarks into a photon.

\begin{figure}[here]
\centering
    \includegraphics[scale=.5]{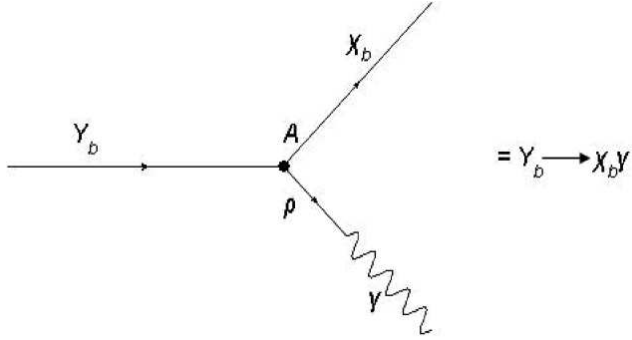}
    \caption{Radiative decay of the $Y(10890)$%
. The radiative decay proceeds through the hadronic transition.}
\end{figure}

Exploiting the Vector Meson Dominance (VMD), one can write
transition matrix from the Fig. 4-1 as:

\[
\langle \chi _{b}\gamma |Y_{b}\rangle =\langle \gamma |\rho \rangle \frac{1}{%
m_{\rho }^{2}}\langle \chi _{b}\rho |Y_{b}\rangle =\frac{f_{\rho
}}{m_{\rho }^{2}}A
\]%
Thus the partial decay width is:%
\[
\Gamma (Y_{b}\rightarrow \chi _{b}\gamma )=2|A^{2}|\left( \frac{f_{\rho }}{%
m_{\rho }^{2}}\right) ^{2}\frac{1}{8\pi
M_{Y_{b}}^{2}}\frac{\sqrt{\lambda (M_{Y_{b}},\ M_{\chi _{b}},\
0)}}{2M_{Y_{b}}}
\]
where $\frac{\sqrt{\lambda (M_{Y_{b}},\ M_{\chi _{b}},\
0)}}{2M_{Y_{b}}}$ is the decay momentum and the value of $\lambda
(M_{Y_{b}},\ M_{\chi _{b}},\ 0)$ is very easy to calculate using
Eq.(\ref{2.48}).

Using $f_{\rho }=0.152$ GeV$^{2}$ given in \cite{21}, we get:%
\[
\frac{\Gamma (Y_{b}\rightarrow \chi _{b}\gamma )}{\Gamma
(Y_{b}\rightarrow \Upsilon (1S)\;\pi ^{+}\pi ^{-})}\sim 0.3
\]%
Similarly%
\[
\frac{\Gamma (Y_{b}\rightarrow \eta _{b}\gamma )}{\Gamma
(Y_{b}\rightarrow \Upsilon (1S)\;\pi ^{+}\pi ^{-})}\sim 0.5
\]%
For $Y_{b}\rightarrow \chi _{b}\gamma $ and $Y_{b}\rightarrow \eta
_{b}\gamma ,$ the lack of experimental results do not allow us to
draw any solid conclusions. However, we have made their theoretical
decay rate predictions.

\chapter{Conclusion}

We have analyzed the spectroscopy of particles with hidden charm and
bottom
in diquark-antidiquark structures of the kind of ${\mathcal{Q}}{\bar{%
\mathcal{Q}}}=[\acute{q}q][\overline{\acute{q}}\bar{q}]$, where
$q=u$ or $d$ ; $\acute{q}=c$ or $b$, and $\mathcal{Q}$ is a diquark.
The idea that the color diquark is handled as a constituent building
block is at the core of the approach taken in this dissertation. In
this context we used the constituent quark model (CQM) as a
benchmark to identify and compare the properties of the newly
discovered. Constituent quark models typically assume a QCD
motivated potential that includes a Coulomb-like one-gluon-exchange
potential at small separation and a linearly confining potential at
large separation. In the constituent quark model hadron masses
are described by an effective Hamiltonian: $H=\sum\limits_{i}m_{i}+\sum%
\limits_{i<j}2\mathcal{K}_{ij}(\mathbf{S}_{i}\cdot \mathbf{S}_{j})$
that takes as input the constituent quark masses and the spin-spin
couplings between quarks. The coefficient $\mathcal{K}_{ij}$ depends
on the flavor of the constituents $i$, $j$ and on the particular
color state of the pair. By extending this approach to
diquark-antidiquark bound states it is possible
to predict tetraquark mass spectra. The mass spectrum of tetraquarks $[%
\acute{q}q][\overline{\acute{q}}\bar{q}]$ with $q=u$, $d$ and
$\acute{q}=c$, $b$ neutral states are described in terms of the
constituent diquark masses, $m_{\mathcal{Q}}$, spin-spin
interactions inside the single diquark, spin-spin interaction
between quark and antiquark belonging to two diquarks, spin-orbit,
and purely orbital term \cite{17}. We calculated the couplings for
color singlet combinations from the known $L_{\mathcal{Q\bar{Q}}}=0$
mesons. Spin-spin coupling for quark-quark in color $\bar{3}$ \
(antitriplet) state are calculated from the known $L=0$ baryons.
According
to one gluon exchange, from Eq.(\ref{1.15}) and Eq.(\ref{1.16}), we have $(%
\mathcal{K}_{ij})_{\bar{3}}=\frac{1}{2}(\mathcal{K}_{ij})_{0}$. This
relation holds for singlet and antitriplet states. The couplings
corresponding to the spin-spin interactions have been calculated for
the color singlet and color antitriplet only. The couplings are not
necessarily in the singlet state but octet couplings $\left( \mathcal{K}_{c%
\bar{c}}\right) _{8}$ are also possible. The quantities $\mathcal{K}_{q\bar{q%
}}$, $\mathcal{K}_{c\bar{q}}$ and $\mathcal{K}_{c\bar{c}}$ involve
both color singlet and color octet couplings between the quarks and
antiquraks in a $\mathcal{Q\bar{Q}}$ system (A quark in the\ diquark
$\mathcal{Q}$ could have a color octet spin-spin interaction with an
antiquark in the
antidiquark $\mathcal{\bar{Q}}$). We have calculated a relationship: $%
\mathcal{K}_{c\bar{c}}\left( [cq][\bar{c}\bar{q}]\right)
=\frac{1}{4}\left( \mathcal{K}_{c\bar{c}}\right) _{0}$, using one
gluon exchange model. The
states are classified in terms of the diquark and antidiquark spin, $S_{%
\mathcal{Q}}$ and $S_{\bar{\mathcal{Q}}}$, total angular momentum
$J$,
parity $P$ and charge conjugation, $C$. We have\ considered both \textit{%
good }($S_{\bar{\mathcal{Q}}}=0$) and \textit{bad} ($S_{\bar{\mathcal{Q}}}=1$%
) diquraks for which $L_{\mathcal{Q\bar{Q}}}=0$, we have six
possible states. By diagonalizing the Hamiltonian given in
Eq.(\ref{2.34}) and using the spin couplings we have calculated the
masses of the tetraquark states. The mass of the diquark $[cq]$ was
fixed by using the mass of $X(3872)$ as input, yielding
$m_{[cq]}=1.933$ $GeV$. The $1^{++}$ state is a good candidate\ to
explain the properties of $X(3872).$ In order to reduce the
experimental information needed we estimate the remaining diquark
masses by substituting the costituent quark forming the diquark.
Isospin-breaking
introduces a mass splitting and the mass eigenstates called $X_{[c,l]}$ and $%
X_{[c,h]}$ (for lighter and heavier of the two) become linear
combinations of $X_{[cu]}$ and $X_{[cd]}$. One can put:
$X_{[c,l]}=\cos \theta \ X_{[cu]}\;+\sin \theta \;X_{[cd]}$ and
$X_{[c,h]}=-\sin \theta \ X_{[cu]}+\cos \theta \ X_{[cd]}.$ Besides
these two neutral states, two
charged states arise as a natural prediction of the tetraquark picture $%
X^{+}=[cu][\bar{c}\bar{d}]$ $and$ $X^{-}=[cd][\bar{c}\bar{u}]$. The
charged partners are $X^{\pm }$ are not observed \cite{30}.

We modified the formulism of diquark-antiquark model to calculate
the
spectrum of hidden bottom (bottomness $=0$) states for $L_{\mathcal{Q\bar{Q}}%
}=1$. We have shown that $Y_{b}$ is $J^{PC}=1^{--}$ state, with $%
Y_{[bq]}=([bq]_{S=0}[\bar{b}\bar{q}]_{S=0})_{\mathrm{P-wave}},$ with
the value $M_{Y_{[bq]}}^{(1)}$ $($for $q=u,d)$ equal to $10890$ MeV.
We identify
this with the mass of the $Y_{b}$ from Belle \cite{37}, apart from the $%
\Upsilon (5S)$ and $\Upsilon (6S)$ resonances.

We discussed the decays modes of $X(3872)$ and $Y(10890)$ states on
the basis of quark rearangement in the
${\mathcal{Q}}{\bar{\mathcal{Q}}}$ system. Originally the $X(3872)$
was found through its decay into $J/\psi \pi ^{+}\pi ^{-}.$ The
decay of a diquark-antidiquark bound state into a pair of mesons can
occur through the exchange of a quark and an antiquark belonging
respectively to the diquark and the antidiquark. The only
available ones are $J/\psi \rightarrow 2\pi $ and $J/\psi \rightarrow 3\pi $%
, dominated by $\rho ^{0}$ and $\omega $, experimentaly confirmed
respectively. We have calculated hadronic decay widths: $\Gamma
(X_{[c,l]}\rightarrow \psi +\pi ^{+}\pi ^{-})=3.78$ $MeV$ and
$\Gamma (X_{[c,l]}\rightarrow \psi +\pi ^{+}\pi ^{-}\pi ^{0})=0.66$
$MeV.$ We got information on the mixing angle from the decay
rates:$\left( \frac{\Gamma (3\pi )}{\Gamma (2\pi )}\right)
_{X_{[c,l]}}=\frac{(\cos \theta \ +\sin \theta )^{2}}{(\cos \theta \
-\sin \theta )^{2}}\cdot \frac{\langle p\rangle _{\omega }}{\langle
p\rangle _{\rho }}=0.802.$ But $\left( \frac{\Gamma (3\pi )}{\Gamma
(2\pi )}\right) _{Belle}=0.8\pm 0.3_{stat}\pm 0.1_{syst}.$ Thus
$\theta \ =\pm 20^{0}$ , for $X_{[c,l]}$ and $X_{[c,h]}$
respectively.
For the charged state $X^{\pm },$ that decay via $\rho $-exchange only: $%
\Gamma (X^{\pm }\rightarrow J/\psi \pi ^{\pm }\pi ^{0})=2|A^{2}|\frac{%
\langle p\rangle _{\rho }}{8\pi M_{X}^{2}}=4.6$ MeV. The amplitude
for the radiative decay $X$ $\rightarrow $ $J/\psi \gamma $ is
calculated using the Vector Meson Dominance (VMD). The radiative
decay proceeds through the hadronic transition $X\rightarrow J/\psi
\rho .$ We found that the ratio of radiative decay to hadronic ones
i-e, $\frac{\Gamma (X\rightarrow J/\psi
\gamma )}{\Gamma (X\rightarrow J/\psi \pi ^{+}\pi ^{-})}\sim 0.44$ and $%
\frac{\Gamma (X\rightarrow \chi _{c1}\gamma )}{\Gamma (X\rightarrow
J/\psi \pi ^{+}\pi ^{-})}\sim 0.89$ that are in agreement with
experimental values
\cite{31}. We exploited the result obtained for the width of $X$ $%
\rightarrow $ $J/\psi \gamma $\ to give an estimate of the decay width into $%
J/\psi \gamma \gamma $. We have obtained $\frac{\Gamma (X\rightarrow
\gamma \gamma )}{\Gamma (X\rightarrow J/\psi \pi ^{+}\pi ^{-})}\sim
3\times 10^{-5}$
that is greater than the upper limit provided in experimental data \cite{54}%
. The inconsistency of the theoretical prediction with respect to
data is not dramatic if one we take into account the very strong
assumptions made to derive the decay widths.

For bottomonium systems, the corresponding decay widths are
determined by
the wave functions at the origin for the $\Upsilon (nS)$, $\Psi _{b\bar{b}%
}(0)$, and by the derivative of these functions at the origin, $\Psi _{b\bar{%
b}}^{\prime }(0)$, for the P-waves. Due to the possibly larger
hadronic size of the tetraquarks compared to that of the $b\bar{b}$
mesons, we modified the Quarkonia potential. For example, the
Buchm\"{u}ller-Tye $Q\bar{Q}$
potential~\cite{38}. This will modify the tetraquark wave functions $\Psi _{%
\mathcal{Q\bar{Q}}}(0)$ from the corresponding wave functions of the bound $b%
\bar{b}$ systems, effecting the decay amplitudes and hence all the
decay widths of the tetraquarks. The corresponding value for the
tetraquark states $[bq][\bar{b}\bar{q}]$ is then calculated taking
into account the ratio of the string tensions $\kappa
=k_{\mathcal{Q\bar{Q}}}/k_{Q\bar{Q}},$ string tension in a diquark
$k_{\mathcal{QQ}}$ is expected to be different than the
corresponding string tension $k_{Q\bar{Q}}$ in the $Q\bar{Q}$
mesons. We
expect $\kappa $ to have a value in the range $\kappa \in \lbrack \frac{1}{2}%
,\frac{\sqrt{3}}{2}].$ As the linear part of the confining potential
determines essentially the heavy Quarkonia wave functions, we find
that to a
good approximation: $\Psi _{\mathcal{Q\bar{Q}}}(0)\simeq \kappa \Psi _{b\bar{%
b}}(0).$ The partial electronic widths for the P-states are given by
the well known Van Royen-Weisskopf formula. For the $Y_{b}(10890)$
state we have obtained $\Gamma _{ee}\approx 0.12KeV$. This value is
close to the experimental value given in \cite{30}. We have also
calculated the two-body hadronic decays of the $Y_{b}(10890)$,
$Y_{b}(q)\rightarrow B_{q}^{(\ast
)}(k)\bar{B}_{q}^{(\ast )}(l).$ The amplitude for the radiative decay $%
Y_{b}\rightarrow $ $\chi _{b}\gamma $ will\ proceed through the
annihilation of a pair of light quarks into a photon. We have
calculated $\frac{\Gamma (Y_{b}\rightarrow \chi _{b}\gamma )}{\Gamma
(Y_{b}\rightarrow \Upsilon (1S)\;\pi ^{+}\pi ^{-})}\sim 0.3$ and
$\frac{\Gamma (Y_{b}\rightarrow \eta _{b}\gamma )}{\Gamma
(Y_{b}\rightarrow \Upsilon (1S)\;\pi ^{+}\pi ^{-})}\sim 0.5$ of
which we do not have any experimental results. We expect that our
predictions should provide guidance for the future.

This work will be of interest both for theoretical and experimental
particle physicists for the next couple of years. Now the next task
is to explore the tetraquark state $[bq][\bar{c}\bar{q}].$ The
motivation is in fact that the
searches for tetraquarks in the $B_{c}$ system may find these $[bq][\bar{c}%
\bar{q}]$ tetraquark below the $B\bar{D}$ threshold. Possible exotic
signature include strong or electromegnetic decays in to $B_{c}\pi $ or $%
B_{c}\gamma $, weak decays producing additional peaks in the mass
spectrum of $B_{c}$ decay final state.

\end{document}